\documentclass[
aps,prd,twocolumn,reprint,amsmath,amssymb,preprintnumbers,superscriptaddress,nobibnotes,nofootinbib,floatfix,longbib
]{revtex4-1}

\usepackage{graphicx}
\usepackage{dcolumn}
\usepackage{bm}
\usepackage{color}
\usepackage{amssymb}
\usepackage{amsmath}
\usepackage{braket}
\usepackage[colorlinks=true,allcolors=Blue,pdfusetitle]{hyperref}
\usepackage{rotating}
\usepackage{multirow}
\usepackage{graphicx}
\usepackage[dvipsnames]{xcolor}
\usepackage{svg}
\usepackage{esdiff}
\usepackage[normalem]{ulem}

\definecolor{Green}{RGB}{0, 128, 0}
\newcommand{\orcid}[1]{\href{https://orcid.org/#1}{\includegraphics[width=10pt]{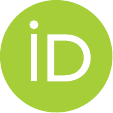}}}

\usepackage{physics}

\newcommand\redsout{\bgroup\markoverwith{\textcolor{red}{\rule[0.5ex]{2pt}{0.4pt}}}\ULon}

\begin{document}
\preprint{LLNL-JRNL-2004830-DRAFT}
\title{Quantum simulations of nuclear resonances with variational methods}
\author{Ashutosh Singh}
\email{ashutosh\_s@ph.iitr.ac.in}
\affiliation{Department of Physics, Indian Institute of Technology Roorkee, Roorkee 247667, Uttarakhand, India}

\author{Pooja Siwach \orcid{0000-0001-6186-0555}}
\email{siwach1@llnl.gov}
\affiliation{Nuclear and Chemical Science Division, Lawrence Livermore National Laboratory, Livermore, California 94551, USA}

\author{P. Arumugam \orcid{0000-0001-9624-8024}}
\email{arumugam@ph.iitr.ac.in}
\affiliation{Department of Physics, Indian Institute of Technology Roorkee, Roorkee 247667, Uttarakhand, India}

\date{\today}

\begin{abstract}
\begin{description} 
\item[Background] The many-body nature of nuclear physics problems poses significant computational challenges. These challenges become even more pronounced when studying the resonance states of nuclear systems, which are governed by the non-Hermitian Hamiltonian. Quantum computing, particularly for quantum many-body systems, offers a promising alternative, especially within the constraints of current noisy intermediate-scale quantum (NISQ) devices.
\item[Purpose] This work aims to simulate nuclear resonances using quantum algorithms by developing a variational framework compatible with non-Hermitian Hamiltonians and implementing it fully on a quantum simulator.
\item[Methods] We employ the complex scaling technique to extract resonance positions classically and adapt it for quantum simulations using a two-step algorithm. First, we transform the non-Hermitian Hamiltonian into a Hermitian form by using the energy variance as a cost function within a variational framework. Second, we perform $\theta$-trajectory calculations to determine optimal resonance positions in the complex energy plane. To address resource constraints on NISQ devices, we utilize Gray Code (GC) encoding to reduce qubit requirements.
\item[Results] We first validate our approach using a schematic potential model that mimics a nuclear potential, successfully reproducing known resonance energies with high fidelity. We then extend the method to a more realistic $\alpha -\alpha$ nuclear potential and compute the D- and G-wave resonance energies with a basis size of $N=16$, using only four qubits. The quantum simulation results closely match the classical values, demonstrating the feasibility of our approach.
\item[Conclusions] This study demonstrates, for the first time, that the complete $\theta$-trajectory method can be implemented on a quantum computer without relying on any classical input beyond the Hamiltonian. The results establish a scalable and efficient quantum framework for simulating resonance phenomena in nuclear systems. This work represents a significant step toward quantum simulations of open quantum systems and lays the foundation for future investigations into resonance structures in nuclear, atomic, and molecular physics.
\end{description}
\end{abstract}

\maketitle

\section{Introduction}
The rapid advancements in quantum algorithms and hardware have significantly propelled the field of quantum computing, particularly in applications related to simulating quantum many-body systems~\cite{RevModPhys.94.015004_9, refId0_martin_18, McClean_2016_19}. A notable application in this domain is the evaluation of a system’s energy spectrum. Among the earliest algorithms proposed for this purpose, quantum phase estimation (QPE) stands out, offering linear scaling compared to classical methods~\cite{PhysRevD.105.074503_4, turro2024evaluationphaseshiftsnonrelativistic_6, PhysRevA.104.012611_scattering_7}. However, QPE is highly resource-intensive, making it impractical for the current generation of noisy intermediate-scale quantum (NISQ) devices.

To harness the capabilities of NISQ devices, hybrid quantum-classical algorithms have emerged as more feasible alternatives~\cite{nature_23, PhysRevLett.120.210501_prl_deuteron_8, variational_npj_21, PhysRevA.108.032417_reaction_3, Yuan2019theoryofvariational_12, PhysRevC.108.064305_chand, PhysRevC.104.024305_lipkin_model_11, Stokes2020quantumnatural_14}. The variational quantum eigensolver (VQE) is a prominent example, designed to compute the ground-state energy of a system by leveraging the variational principle of quantum mechanics. VQE has been successfully employed to simulate a variety of systems in both physics and quantum chemistry. Moreover, advanced adaptations of VQE have been developed to enhance its efficiency and broaden its applicability~\cite{Higgott2019variationalquantum_vqd_10, PhysRevResearch.2.043140_molecular_excitation_13, PRXQuantum.2.020310_qubit_adapt_vqe_15, PhysRevC.105.064317_adapt_vqe_16, adapt_nature_22}.

However, the aforementioned quantum algorithms generally assume the systems under consideration to be closed, resulting in Hermitian Hamiltonians and real eigenvalues. In such cases, the eigenvectors are orthogonal to one another. Yet, many natural phenomena are governed by non-Hermitian quantum mechanics, where eigenvalues are not always real. This non-Hermiticity poses a challenge for VQE, as the algorithm relies on energy minimization, which is not well-defined in this context. To address this issue, the energy variance can be used as an alternative cost function. For an eigenstate, the energy variance is zero, providing a viable metric for optimization. A variational method tailored to compute resonances in non-Hermitian systems leverages this principle~\cite{xie:fp2024}, enabling the study of resonances.

Several alternative algorithms have been proposed for simulating resonances on quantum computers. One such approach is the direct measurement method based on the linear combination of unitaries (LCU) algorithm, which has been used to calculate the energy of resonant states in atomic and molecular systems~\cite{Bian:jcp2021} and nuclear physics~\cite{zhang:plb2024}. This method works by decomposing the non-unitary Hamiltonian into a combination of unitary operators. However, its applicability is restricted to very small systems, typically involving only two qubits, due to the exponential growth in resource requirements with increasing system size.

Another variational method, called the variational quantum universal eigensolver (VQUE), was proposed in Ref.~\cite{Zhao:sr2023} to compute the eigenvalues of non-Hermitian matrices based on Schur’s triangularization theory. In VQUE, a non-Hermitian matrix is triangularized using a unitary matrix, which is determined iteratively through variational quantum algorithms. Additionally, non-unitary gate operations are implemented using the LCU method. As a result, VQUE is also highly resource-intensive, making it challenging to execute on NISQ devices.

In this work, we aim to simulate resonances in quantum many-body systems. Resonance states are governed by non-Hermitian Hamiltonians, characterized by complex eigenvalues, and are a fundamental feature of numerous phenomena in physics and beyond. These resonances are quasistationary states that lie between bound and continuum states. Their wave functions do not exhibit normal decay properties as $r\to \infty$; instead, they grow exponentially in this limit, rendering them non-square-integrable. To address this challenge, similarity transformations, such as complex scaling, can be employed. This approach transforms the problem of locating resonance positions on the complex energy plane into an eigenvalue problem for a non-Hermitian operator in Hilbert space. In this work, we use the complex scaling method (CSM) to compute resonance states on a quantum computer. In CSM, the Hamiltonian is rotated by an angle $\theta$, which results in square-integrable wave functions. This method enables the treatment of bound and resonance states on the same footing.

To perform quantum simulations of resonances, we use the variational method proposed in Ref.~\cite{Zhao:sr2023}, which is based on the variance of energy. Where, the spectrum of complex eigenvalues is scanned by initializing the energy parameter close to the possible eigenvalue. However, this approach is impractical, as it requires prior knowledge of the spectrum or necessitates rerunning the algorithm multiple times. In this work, we propose a more efficient strategy that utilizes the properties of eigenvalues with respect to the rotation angle $\theta$, enabling us to scan the spectrum, and extract optimal resonance position without relying on prior knowledge of resonance positions. To validate our methods, we utilize a schematic model from Ref.~\cite{myo:2020ptepptaa101}, which closely resembles a nuclear potential with a barrier. Additionally, we simulate resonances in nucleus-nucleus scattering using the complex scaling method, achieving results with significantly improved accuracy compared to previous studies~\cite{zhang:plb2024}.

Calculating resonance states requires a large basis size, necessitating a significant number of qubits, which poses challenges for quantum computation. To address this, we employ GC encoding to reduce the number of qubits and ADAPT-VQE to minimize the quantum gate count, thereby achieving accurate results with quantum computers.

The rotation angle $\theta$ in CSM is determined by analyzing the stationary points along the trajectory of complex eigenvalues. While the CSM has been previously applied to study resonances in atomic and nuclear systems (Refs.~\cite{Bian:jcp2021,zhang:plb2024}), the angle $\theta$ in these studies was fixed apriori. In this work, we propose a novel method to evaluate $\theta$ dynamically on a quantum computer.

This paper is organized as follows: In Sec.~\ref{sec:comp_res}, we discuss the complex scaling method and the optimal choice of basis set for this approach. Sec.~\ref{sec:model_Hamiltonian} and Sec.~\ref{sec:qubit_mapping} present the model Hamiltonian used in our calculations and the procedure for mapping it to qubit space, respectively. In Sec.~\ref{sec:quantum_algorithm}, we describe the algorithms used to extract the optimal resonance positions and devise a mechanism to sort non-physical states. In Sec.~\ref{sec:result_disscuss}, we present and discuss our results for two model potentials used to describe nuclear resonances. Finally, we conclude our work in Sec.~\ref{sec:conclussion}.

\section{Computation of resonances}\label{sec:comp_res}
\begin{figure*}
    \centering
    \includegraphics[width=.95\linewidth]{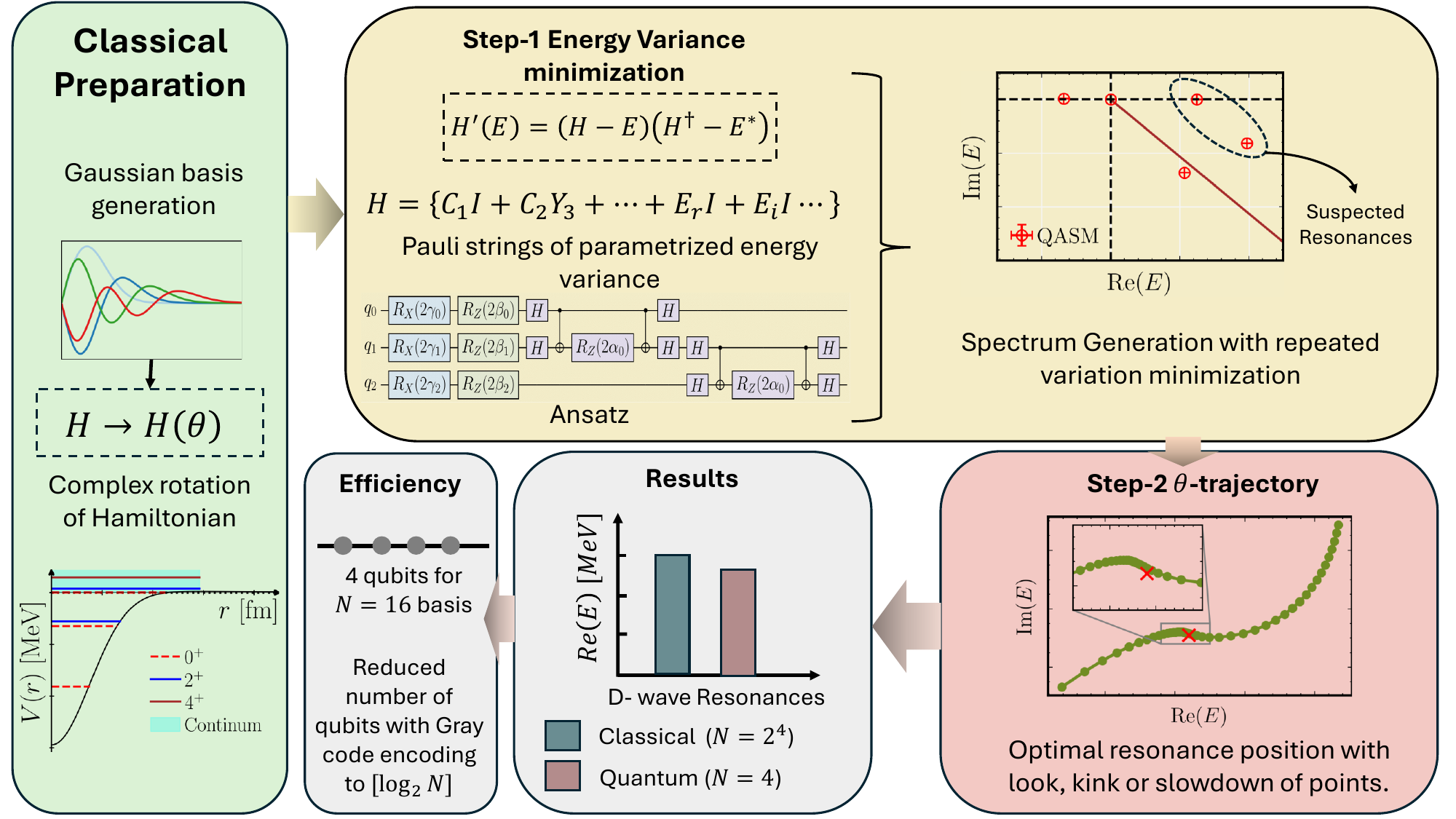}
    \caption{Schematic workflow illustrating the procedure for extracting optimal resonance positions using a quantum computer. The process begins with Hamiltonian construction on a classical computer, followed by the mapping of basis functions and the Hamiltonian to qubits and Pauli strings. A parameterized energy variance minimization is then used to calculate the complex energy eigenvalues, where energy is treated as a variational parameter. By repeating this procedure with incrementally varied initial energy guesses, a complete spectrum is generated, and suspected resonances—identified by their separation from the continuum in the lower complex energy plane—are located. A $\theta$-trajectory calculation is subsequently performed to extract the optimal resonance position, indicated by a loop, kink, or slowdown along the trajectory. The overall efficiency of the method is enhanced by employing GC encoding, which allows the mapping of a basis size of $N=16$ to just four qubits.}
    \label{fig:workflow_diagram}
\end{figure*}

Quantum scattering theory has been widely used to study resonances. The real and imaginary parts of the complex energy correspond to the energy and width of the resonance, respectively, and are determined as the poles of the S-matrix. However, scattering theory falls short in describing many-body resonances. Moreover, these poles, except for those representing bound states, are associated with exponentially divergent wave functions. This divergence presents significant challenges in the framework of non-Hermitian quantum mechanics, particularly in defining the inner product, which is crucial for applying quantum computational methods. To address this, the primary objective is to make these wave functions square-integrable by applying specific mathematical transformations, such as the complex scaling method. Before moving on to the complex scaling method, we present a schematic workflow outlining the key steps involved in extracting nuclear resonance positions using quantum computation is shown in Fig.~\ref{fig:workflow_diagram}.

\subsection{Complex scaling method}
In the complex scaling method, the coordinates are rotated in the complex plane along the direction of divergence. This transformation renders divergent wave functions square-integrable, allowing the evaluation of observables using standard quantum mechanical tools. Specifically, the coordinates are scaled by a complex phase factor, $x \to xe^{i\theta}$, with analogous transformations applied to the $y$- and $z$-coordinates. The transformation operator $U(\theta)$ transforms the radial $r$ and momentum $k$ co-ordinates as follows
\begin{equation}
    U(\theta)\vb{r}U^{-1}(\theta)=\vb{r}e^{i\theta}, \hspace{5pt} U(\theta)\vb{k}U^{-1}(\theta)=\vb{k}e^{-i\theta}.
\end{equation}
Here, $U(\theta)$ is a unitary operator satisfying $U(\theta)U^{-1}(\theta)=1$.

The single-particle Hamiltonian is given by
\begin{equation}
    H = T +V, \label{eq:hamiltonian}
\end{equation}
where $T$ and $V$ are the kinetic and potential energy terms, respectively.
The transformed Hamiltonian under operator $U(\theta)$ takes the following form
\begin{equation}
    H(\vb{r}, \theta) = U(\theta)H(\vb{r})U(\theta)^{-1},  \label{eq:hamiltonian_similarity_trans}
\end{equation}
and the corresponding wave function is given by
\begin{equation}
    \Psi(\vb{r}, \theta) =  U(\theta)\Psi\qty(\vb{r}).   \label{eq:wave_fn_trans}
\end{equation}
Using the Eqs.~\eqref{eq:hamiltonian_similarity_trans} and~\eqref{eq:wave_fn_trans},  Schr\"odinger equation transforms to
\begin{equation}
    H(\vb{r}, \theta) \Psi(\vb{r}, \theta) = E(\theta)\Psi(\vb{r}, \theta),
\end{equation}
and the Hamiltonian given in Eq.~\eqref{eq:hamiltonian} transforms to
\begin{equation}
    H(\vb{r},\theta) = \exp(-2i\theta)T + V\qty(\vb{r}e^{i\theta}). \label{eq:comp_rot_ham}
\end{equation}

\begin{figure}
    \centering
    \includegraphics[width=0.7\linewidth]{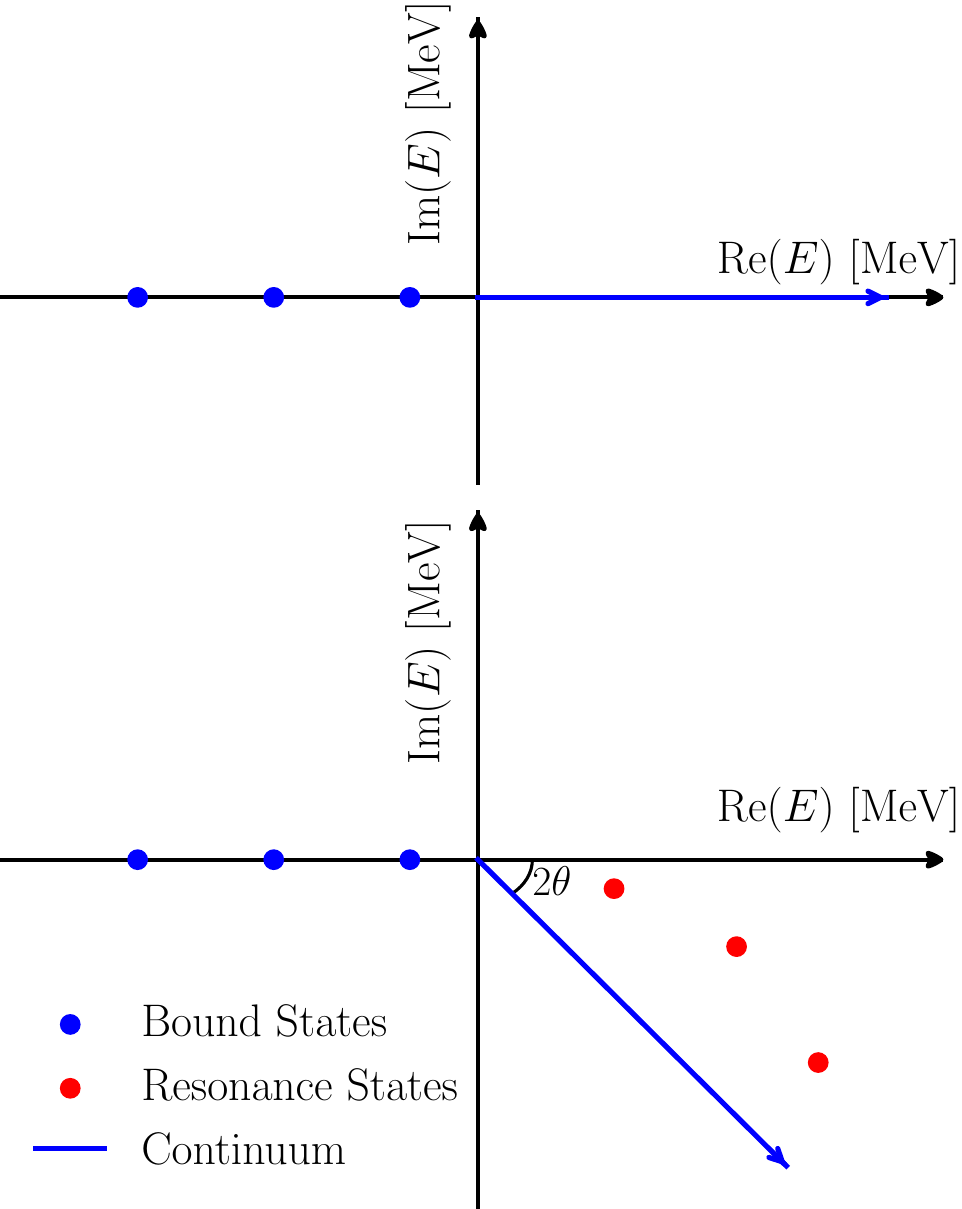}
    \caption{The spectra of typical Hamiltonian without (top) and with (bottom) rotation. Blue dots represent the bound states on a negative part of the horizontal axis, red dots represent the resonant states uncovered after complex rotation, and the blue lines represent the continuum.}
    \label{fig:complex_rotated_spectra}
\end{figure}

It is proven by Aguilar, Balslev, and Combes, know as the ABC theorem~\cite{abc:Aguilar, abc:Balsev}, that the continuum of $H(\vb{r},\theta)$ is rotated down by $2\theta$ in the lower half of the complex energy plane, while discrete resonance energies remain at their original positions as shown in Fig.~\ref{fig:complex_rotated_spectra}. In CSM, complex energy becomes independent of the scaling parameter $\theta$, if $\theta > \theta_{c} = \frac{1}{2}\tan^{-1}(\Gamma / 2 E_r)$, where $\theta_c$ is critical angle, $\Gamma = 2E_i$ is decay width, $E_r$ and $E_i$ are real and imaginary part of energy respectively. However, in realistic calculations due to the finite basis, it is not completely independent of scaling parameter $\theta$.

To calculate the eigenvalues of the transformed Hamiltonian, we expand the rotated wave function $\Psi(\vb{r}, \theta)$ in a basis set of real square-integrable function $\qty{\phi_{i}(\vb{r}): i = 1, 2,...,N}$ such as
\begin{equation}\label{eq:basis_expansion}
    \Psi(\vb{r}, \theta) = \sum_{i=1}^{N}c_{i}(\theta)\phi_{i}(\vb{r}).  
\end{equation}
The coefficients $c_{i}(\theta)$ can be obtained by generalized variational principle~\cite{Moiseyev:PhysRevC.37.383} defined as
\begin{equation}
    \delta \qty[\frac{\int dr \Bar{\Psi}_{N}(\vb{r},\theta)^{*}H(\vb{r},\theta)\Psi_{N}(\vb{r},\theta)}{\int dr \Bar{\Psi}_{N}(\vb{r},\theta)^{*}\Psi_{N}(\vb{r},\theta)}] = 0.
\end{equation}
Here, the bar on the functions means that only their angular factors are to be complex conjugated. The above equation results in a generalized eigenvalue equation~\cite{2011nhqm.book.....M,d_kraft_2013_thessis} given by
\begin{equation}\label{eq:gen_eig_val_eq}
    \sum_{j=1}^{N}\qty[H_{ij}(\theta) - E(\theta)N_{ij}]c_{j}(\theta) = 0,  
\end{equation}
where 
\begin{align}
    H_{ij} =&  \int dr \Bar{\phi}_i(\vb{r})^{*}H(\vb{r},\theta)\phi_i(\vb{r}), \\
    N_{ij} =& \int dr \Bar{\phi}_i(\vb{r})^{*}\phi_i(\vb{r}).
\end{align}
The best estimate for resonance energy is obtained at a specific $\theta$, where the rate of change of energy with respect to $\theta $ is minimal, {\it i.e.}, $\frac{\dd{E(\theta)}}{\dd{\theta}} = 0$. To determine the optimal $\theta$, we plot the $\theta$-trajectory for a given basis size $N$. The optimal resonance position corresponds to a point on the complex energy plane where the trajectory forms a loop or a kink, or where the rate of change in the energy position significantly slows down.
\subsection{Optimal basis sets} \label{subsec:optimal_basis}
To compute the eigenvalues of the Hamiltonian, the harmonic oscillator (HO) basis and the Gaussian basis are among the most widely used. To choose a suitable basis set for calculations, one must determine which basis set better suits the problem of our interest or provides more accurate results. The HO basis is orthonormal, while the Gaussian basis is not. However, Gaussian basis functions can be orthogonalized using the Gram-Schmidt procedure. An effective variational basis set should accurately capture both the rapid changes in the wave function due to short-range correlations and the long-range asymptotic behavior \cite{HIYAMA:ppnp2003}, which the Gaussian basis functions achieve effectively.
Additionally, as suggested by A. Csótó et al.~\cite{csoto:pra1990}, the quality of the approximation in the back-rotated wave function, which is used to recover the original wave function, strongly depends on the chosen basis set. For any well-behaved potential, the Schrödinger equation after complex scaling retains bound state eigenvalues and reveals resonances in the complex energy plane. The eigenfunction $\Psi(\vb{r},\theta)$ obtained in this framework is square-integrable and can be determined using any basis set. The original wave function is recovered by back rotation: $\Psi(r) = U(\theta)^{-1}\Psi(\vb{r},\theta)$. However, back rotation often introduces unwanted oscillations in the solution, regardless of the accuracy of $\Psi(\vb{r},\theta)$.
Notably, the HO basis produces rapid, large-amplitude oscillations in the back-rotated wave function, irrespective of the method used to obtain $\Psi(r)$. In contrast, these oscillations are less pronounced when using the Gaussian basis. Thus, based on these considerations, the Gaussian basis set is determined to be the more suitable choice for the schematic potential in the present work, while the HO basis is better suited for the local potential used to describe nucleus-nucleus scattering.

The HO basis is represented by $\Phi_{nl}(r)$ as
\begin{equation}
    \Phi_{nlm}(\vb{r}) = \Phi_{nl}(r)Y_{lm}(\vartheta, \phi),
\end{equation}
where $n$, $l$, and $m$ are the principal quantum number, the orbital angular momentum quantum number, and its projection quantum number, respectively. $\Phi_{nl}(r)$ is the radial part of the wave function given by
\begin{equation}\label{eq:HO_basis}
    \Phi_{nl}(r) = \qty[\frac{2\Gamma(n+1)}{\Gamma(n+l+\frac{3}{2})}]^{\frac{1}{2}}\frac{1}{b^{\frac{3}{2}}}\qty(\frac{r}{b})^{l}\exp(\frac{-r^2}{2b^2})L_{n}^{l+\frac{1}{2}}\qty(\frac{r^2}{b^2}).
\end{equation}
Here, $L_{n}^{l+\frac{1}{2}}$ denotes the associated Laguerre polynomial, $b$ is the oscillator length parameter, and $n$ takes values $n=0,1,2,\dots n_{\text{max}}$, and $Y_{lm}(\vartheta, \phi)$ are the spherical harmonics.

Gaussian basis functions are defined as
\begin{equation}
    \Phi_{nlm}(\vb{r}) = \Phi_{nl}(r)Y_{lm}(\vartheta, \phi),
\end{equation}
where $\Phi_{nl}(r)$ is the radial part of the wave function given by
\begin{equation}\label{eq:Gauss_bas}
\Phi_{nl}(r) = r^l\qty[\frac{2(2\alpha_n)^{l + \frac{3}{2}}}{\Gamma(l + \frac{3}{2})}]^{\frac{1}{2}}\exp(-\alpha_n r^2).
\end{equation}
Here, $n = 0,1,2,\dots,n_{\text{max}}$, and the parameter $\alpha_n$ is defined as
\begin{align} \alpha_n = \frac{1}{r_n^2}, \notag \ r_n = r_1 a^{n-1}, \label{eq:patameter_gp} \end{align}
where $r_1$ and $a$ are the size parameters, and these geometric size parameters are considered to be the best Gaussian size parameters~\cite{HIYAMA:ppnp2003,csoto:pra1990}. Thus, three parameters can be chosen as $\qty{n_{\text{max}}, r_1, a}$ or $\qty{n_{\text{max}}, r_1, r_{n_{\text{max}}}}$ and in this work, we have considered the later one. The angular part of the wave function is described by the spherical harmonics $Y_{lm}(\vartheta, \phi)$.

We use the Gaussian basis to expand the wave function described in Eq.~\eqref{eq:basis_expansion} to compute the Hamiltonian matrix $H$ and overlap matrix $N$. These matrices are then used to solve the generalized eigenvalue equation, as shown in Eq.~\eqref{eq:gen_eig_val_eq}. The resulting resonance eigenvalue $E(\theta)$ typically depends on three parameters: $r_1$, $\theta$, and $N$ \cite{Moiseyev:mole1978}, 
\begin{equation}\label{eq:e_theta}
    E = E(r_1, \theta, N),
\end{equation}
where $\theta$ is the complex scaling angle, $r_1$ is the non-linear scaling parameter in the Gaussian basis set, and $N$ is the size of the basis set used.

When the non-linear scaling parameter $r_1$ and the rotation angle $\theta$ are fixed, increasing the basis size $N$ causes the resonance eigenvalue to spiral inward toward a limiting position, which represents the best estimate of the true resonance location. To reduce computational complexity, however, we aim to use a smaller basis size. In our approach, we fix $r_1$ and $N$ and vary the rotation angle $\theta$ until the resonance eigenvalue stabilizes along its trajectory, providing a reliable estimate of the true resonance position. As noted earlier, the trajectory of $E$ with $\theta$ often exhibits loops, kinks, or bends as it converges toward the true resonance position.

To solve the problem on a quantum computer using the variational principle, we require a Hamiltonian matrix constructed from a proper orthonormal basis set. While the Gaussian basis set $\{\Phi_{nl}({r}): n=1,\ldots,n_{\text{max}}\}$ is non-orthogonal, we employ the Gram-Schmidt process to transform it into an orthonormal basis suitable for calculating the Hamiltonian matrix on the quantum computer.

\section{Model Hamiltonian}\label{sec:model_Hamiltonian}
We perform quantum simulations of resonances using the CSM with two distinct potentials, a schematic potential serving as a simplified model to demonstrate the application of the technique using the Gaussian basis, and the potential employed to describe nucleus-nucleus interactions, particularly $\alpha-\alpha$ scattering.
\subsection{Schematic potential}
First, we consider a schematic potential to demonstrate the application of quantum computing in simulating resonances using the CSM, expanding the wave functions on a Gaussian basis. The total Hamiltonian of the system can be expressed as
\begin{equation}
    H = -\frac{1}{2}\nabla^2+V(r),
\end{equation}
where $V(r)$ is the schematic potential~\cite{myo:2020ptepptaa101} given by 
\begin{equation}\label{eq:schematic_potential_numerical}
    V(r) = -8\exp(-0.16 r^2) + 4\exp(-0.04 r^2). 
\end{equation}

\begin{figure}
    \centering
    \includegraphics[width=0.9\columnwidth]{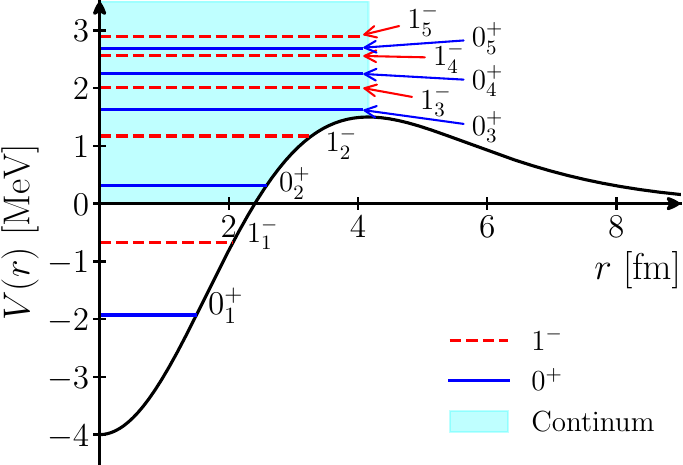}
    \caption{The schematic potential, as defined by Eq.~\eqref{eq:schematic_potential_numerical}, is shown along with several resonances with angular momenta $0^+$ and $1^-$. The plot displays only the real part of the complex energy, which corresponds to the energy, while the imaginary part, representing the decay width, is omitted. The cyan-shaded region represents the continuum, which is discrete in practical calculations due to the use of a finite basis size.}
    \label{fig:schemtaic_potential_complex}
\end{figure}

We calculate the energies of $J=0^+$ and 1$^-$ states. For both cases, there is one bound state and several resonance states. The second states of 0$^+$ and 1$^-$ are quasi-bound states, as they lie below the potential barrier and have very small decay widths. The resonance states just above the potential barrier exhibit maximum resonance energy and are referred to as complex thresholds by Rittby et al. \cite{Rittby:PhysRevA.24.1636}. The shape of the potential and the calculated energies of the bound and resonant states are shown in Fig.~\ref{fig:schemtaic_potential_complex}.

\subsection{Nucleus-nucleus scattering} \label{subsec:nucleus_nucleus_pot}
We further apply the complex scaling method to calculate the resonance states of $\alpha-\alpha$ scattering. The local potential~\cite{Buck:npa1977} for this system is defined as
\begin{eqnarray}
    V_{\alpha\alpha}(r) = V_N(r)+V_c(r),   \label{eq:nucleus-nucleus_pot}
\end{eqnarray}
where $V_N(r)$ is the nucleus-nucleus potential, represented in a local Gaussian form as
\begin{equation}
    V_{N}(r) = V_0\exp(-kr^2),
\end{equation}
and $V_c(r)$ is the Coulomb potential, given by
\begin{equation}
    V_c(r)=\frac{Z_1Z_2e^2}{r}{\rm erf}(\beta r).
\end{equation}
Here, $Z_1$ and $Z_2$ are the atomic numbers of two nuclei ($^4$He in the present case) and $e$ represents the elementary charge. The term ${\rm erf}$ denotes the \emph{error function}. The parameters are chosen as $V_0=-122.6225$ MeV, $k=0.22$ fm$^{-2}$ and $\beta=0.75$ fm$^{-1}$ to reproduce the $0^{+}$ resonance at $92.12\pm0.05$ keV. The measured width of this resonance is $6.8\pm1.7$ eV.

This potential results in three redundant bound states, two $0^+$ states and one $2^+$ state (D-wave), with energies of $-72.7$ MeV, $-25.8$ MeV and $-22.2$ MeV, respectively. The first $2^+$ resonance state is located at $2.88-i0.62$ MeV, while the first $4^+$ (G-wave) resonance state is found at $11.78-i1.78$ MeV.

We employ the HO basis to calculate the resonances of $\alpha-\alpha$ scattering using the CSM.
\begin{figure}
    \centering
    \includegraphics[width=0.9\columnwidth]{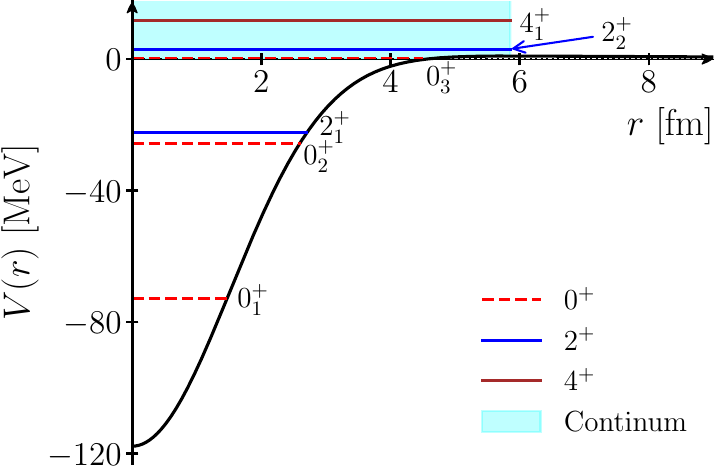}
    \caption{Similar to Fig.~\ref{fig:schemtaic_potential_complex}, the $\alpha-\alpha$ scattering potential, as defined by Eq.~\eqref{eq:nucleus-nucleus_pot}, is shown along with a resonance of $0^+$, $2^+$, and $4^+$ states each.}
    \label{fig:alp_alp_potential_complex_scaling}
\end{figure}
\section{Qubit Mapping} \label{sec:qubit_mapping}
Given a proper orthonormal basis set \{$\phi_{nl}(r)$\}, the Hamiltonian in second quantization can be expressed as
\begin{equation}
H=\sum_{ij}h_{ij}a_{i}^{\dagger}a_{j}, \label{eq:second_quant_ham}
\end{equation}
where
\begin{equation}
    h_{ij}=\bra{i}\hat{T}+\hat{V}\ket{j} =  \int dr \phi_i(r)^{*}H(r)\phi_i(r), \label{eq:hij_mat_ele}
\end{equation}
with $\ket{i}\equiv \phi_{nl}(r)$ representing the basis in wave function notation, and $a_i^{\dagger}$ and $a_i$ being the creation and annihilation operators, respectively. For simplicity, we consider only the one-body term in the second-quantization representation of the Hamiltonian, and $h_{ij}$ is computed on a classical computer. Eq.~\eqref{eq:second_quant_ham} can then be rewritten in terms of Pauli strings by mapping the basis states $\ket{i}$ to qubits using encoding schemes such as one-hot (OH), Bravyi–Kitaev~\cite{10.1063/1.4768229_bk_chem, BRAVYI2002210_bk}, parity or GC~\cite{improve_gc_olivia, pooja_PhysRevC.104.034301_gc}. The creation and annihilation operators $a_{i}^{\dagger}$ and $a_{j}$ are transformed into Pauli gates using transformations like the Jordan-Wigner (JW)~\cite{jw}. In this work, we employ OH and GC mappings to encode our systems on qubits and transform the operators in Pauli strings using the corresponding transformations given in next section.

Following these transformations, the Hamiltonian in terms of Pauli operators is given by
\begin{equation}
    H = \sum_{i=0}^{}c_{i}P_{i},
\end{equation}
where $c_{i}$ represents a coefficient, and $P_{i}$ is a Pauli string represented by a $k$-local tensor product of Pauli operators, where $k \leq n$ and $n$ is the basis size. The number of terms $N_{P}$ in the operator $P$ satisfies $N_{P} < 2^{n}$.

\section{Quantum algorithms} \label{sec:quantum_algorithm}
Since the energies of a non-Hermitian Hamiltonian are not always real, directly minimizing the energy, as done in the VQE, is not practical. However, the variance of energy, which becomes zero when the system is in an eigenstate, can be used as a cost function to find complex eigenvalues. To simulate resonances on a quantum computer, we employ a variational method based on this principle, as proposed in Ref.~\cite{Zhao:sr2023}.

As a first step, a Hermitian Hamiltonian $\mathcal{H}(E)$ is constructed from the non-Hermitian Hamiltonian $H$ as follows
\begin{equation}
    \mathcal{H}(E)=(H^\dagger-E^*)(H-E),
\end{equation}
where $E$ is the energy and $E^*$ is its complex conjugate. In the case of a non-Hermitian Hamiltonian $(H\ne H^\dagger)$, the eigenstates of $H$ and $H^\dagger$ are distinct and are referred to as the right and the left eigenvectors, respectively.

The expectation value of $\mathcal{H}(E)$ with respect to any wave function is always greater than or equal to zero. Therefore, the parameterized energy $E$ and the parameters of the trial wave function $\ket{\psi}$ can be optimized such that
\begin{eqnarray}
    (H-E)\ket{\psi}=0,
\end{eqnarray}
yielding the right eigenvector. To determine the left eigenvector, another Hermitian matrix $\mathcal{H}'(E)$ is defined as
\begin{equation}
    \mathcal{H}'(E)=(H-E)(H^\dagger-E^*).
\end{equation}
Minimizing the expectation value of  $\mathcal{H}'(E)$ provides the left eigenvector.
\subsection{Variational algorithm for nuclear resonances}\label{subsec:var_alg_quant}
In this work, we devise an algorithm to simulate $\theta$-trajectory on a quantum computer purely and to obtain the optimal value of resonance position on a complex energy plane for the schematic nuclear potential. For sort-hand notation we will call it variational quantum algorithm (VQA).


\textbf{Algorithm $\bm{1}$}: Steps involved for quantum simulation:
\begin{enumerate} 
\item Create Gaussian basis states $\Phi(r, r_1,a)$. Orthonormalize the basis set using the Gram-Schmidt procedure.
\item Calculate $h_{ij}(r_{1i},a_{1i},r_{1j},a_{1j}, \theta)$ given in Eq.~\eqref{eq:hij_mat_ele}.
\item Construct the Hermitian Hamiltonian from the non-Hermitian one by Hermitianization technique, given by $\mathcal{H}(E) = ({H}^{\dagger} - E^*)({H} - E)$ with $E = E_r + i E_i$, where $E_r$ and $E_i$ are real and imaginary part of energy.
\item Use a parameterized unitary circuit as the \emph{Ans\"atze}, which is composed of a series of single-qubit and double-qubit rotation gates, and is given by 
\begin{eqnarray}
    U(\zeta) &=& \prod_{j=1}^{P}U_{j}(\zeta_{j}), \label{eq:ansatz}
    \\
    U_{j}(\zeta_{j}) &=& e^{-i H_{xx}(\beta_j)} \nonumber\\
    &~&\times e^{-i H_{z}(\gamma_j)} e^{-i H_{x}(\Delta_j)} ,\label{eq:ansatz_single_block}
\end{eqnarray}

where $H_{xx}(\beta_j) = \sum_{l}\beta_{j,l}X_{l}X_{l+1}, H_{z}(\gamma_j) = \sum_{l}\gamma_{j,l}Z_{l}, H_{x}(\Delta_j) = \sum_{l}\Delta_{j,l}X_{l}$, and $\zeta_{j} = (\beta_j, \gamma_j, \Delta_j)$ is the set of parameter for \emph{Ans\"atze}.
\item Construct the energy parameterized cost function for the complex rotated Hamiltonian given by 
\begin{equation}
    \mathcal{L}(\zeta, E) = \bra{\psi(\zeta)}{({H}^{\dagger} - E^*)({H} - E)}\ket{\psi(\zeta)},
\end{equation}
where $\ket{\psi(\zeta)} = U(\zeta)\ket{0}.$
\item Complex rotated Hamiltonian is then mapped to qubits using OH or GC schemes, and the energy parametrized cost function is calculated in terms of Pauli string.
\item Implement the VQE algorithm for minimizing the cost function with an initial guess for $E_r$ and $E_i$. The optimized value of the energy $E = E_r + iE_i$ and $\zeta$ parameters will give the complex energy eigenvalue and right eigenvector, respectively.
\item Repeat step 7 until one generates a complete or desired spectrum is achieved.
\end{enumerate}

\subsection{$\theta$-trajectory}\label{subsec:theta_traj_quant}
Implementing Algorithm $1$ will give us a complete spectrum for a given complex-scaled parameter $\theta$. However, it is well established that the continuum is rotated by an angle $2\theta$ after complex scaling. Thus, we are only interested in the values that lie away from the continuum in the lower half of the complex energy plane. These are the energy eigenvalues that correspond to resonances. However, this resonance position depends on $r_1, \theta$, and basis size $N$ as given in Eq.~\eqref{eq:e_theta}. Thus, we fix $r_1$ and $N$ to compute the $\theta$-trajectory and hence determine the optimal value of resonance on a quantum computer.

\textbf{Algorithm $\bm{2}$}: To generate complete $\theta$-trajectory on quantum computer:
\begin{enumerate}
    \item Fix the value of $r_1$ and basis size $N$.
    \item Choose the initial guess for $E = E_r + iE_i$ in the close neighborhood of chosen approximate resonance position from Algorithm $1$.
    \item Repeat Algorithm $1$ for different value of $\theta$ in a small increasing step size.
    \item Plot the histogram of the real and the imaginary parts of energy along with $\theta$-trajectory.
    \item Bin with the highest count of the real and the imaginary parts will give us the optimal value of real and imaginary parts of complex energy separately.
\end{enumerate}
\subsection{State filtration}\label{subsec:state_fileteration}
When mapping fermionic states to qubits, the qubit Hilbert space often contains a larger number of basis states than needed to simulate the system. Consequently, redundant states are generated during quantum simulations. While these states do not pose significant issues when calculating the ground state, since the ground state corresponds to the minimum energy, they become problematic when targeting excited states, such as resonances. To locate the true resonances, it is necessary to filter out the redundant states. This can be achieved by leveraging the system's symmetries. For instance, if the eigenstates of the system conserve particle number or spin, states that do not exhibit these symmetries can be excluded. Such filtration can be effectively achieved using the techniques based on the quantum phase estimation (QPE) algorithm proposed in Refs.~\cite{Lacroix:prl2020,Siwach:pra2021}.

Consider a wave function obtained after applying the VQE, expressed as $\ket{\psi}_f=\sum_i c_i\ket{\phi_i}$, which is mapped to $n_q$ qubits. When QPE is applied with a unitary operator $U$ whose eigenvalues are $\exp(i2\pi\theta_i)$, with control operation on $n_r$ number of ancilla qubits, the final wave function becomes
\begin{equation}
    \ket{\psi}_f=\sum_i c_i\ket{\theta_i 2^{n_r}}\otimes\ket{\phi_i}.
\end{equation}
After repeated measurements, the wave function collapses into one of the possible states $\ket{\theta_i 2^{n_r}}\otimes\ket{\phi_i}$ with a probability $|c_i|^2$, where $\theta_i 2^{n_r}$ is a binary string.

For an operator $S$ with known eigenvalues, a corresponding unitary operator $U_S=e^{i\vartheta S}$ can be defined such that its application in QPE projects onto the eigenstates of $S$. This approach can also be employed to verify whether a symmetry is broken in a given state. For instance, in the present work, we consider a single-particle distributed across $N$ basis states. Consequently, the valid states after applying the VQE should have exactly one qubit in the occupied state. To eliminate states that do not satisfy this single-particle criterion, we define a unitary operator $U_N$ derived from the number operator $\mathcal{N}$, with a phase $\vartheta$ determined based on the number of ancilla qubits. The number operator, expressed in terms of Pauli matrices, is given by
\begin{equation}
   \Hat{\mathcal{N}}=\sum_j a_j^\dagger a_j=\frac{1}{2}\sum_j\left(I_j-Z_j\right).
\end{equation}
The corresponding unitary operator $U_N$ is defined as
\begin{equation}
    U_N=\exp\left(i2\pi\frac{\Hat{\mathcal{N}}}{2^{n_r}}\right).
\end{equation}
In expanded form, $U_N$ acts as
\begin{equation}
    U_N=\prod_j \begin{bmatrix}
        1 & 0\\
        0 & e^{i\pi/2^{n_r-1}}\\
    \end{bmatrix}.
\end{equation}
This operator encodes the particle number information in the quantum system, enabling its extraction through quantum measurement.
\section{Results and Discussion} \label{sec:result_disscuss}
We use the complex scaling method to first calculate the optimal resonance position for the schematic potential and the nucleus-nucleus scattering potential discussed in Sec.~\ref{sec:model_Hamiltonian} using classical methods. Classical calculations are performed for a fixed set of parameters across different basis sizes $N$ to determine an optimal basis size for quantum calculations. Given the noisy and limited qubit availability in current quantum computers, our approach focuses on identifying the smallest basis size that still provides an accurate resonance position. Once an appropriate basis size is selected, we proceed with quantum simulations, first generating the complete energy spectrum to locate the resonance states. Finally, we employ a $\theta$-trajectory analysis to refine the optimal resonance position using a quantum simulator.

\subsection{Resonance states using generalized variational principle}
In this section, we perform classical calculations for the potential models described in Sec.~\ref{sec:model_Hamiltonian} using various basis sizes. We present the $\theta$-trajectories to extract optimal resonance positions, and finally, we identify the optimal basis size and parameters suitable for quantum simulations.
\subsubsection{Schematic potential} \label{subsubsec:scematic_discu_classical}
\begin{figure}
    \centering
    \includegraphics[width=0.9\linewidth]{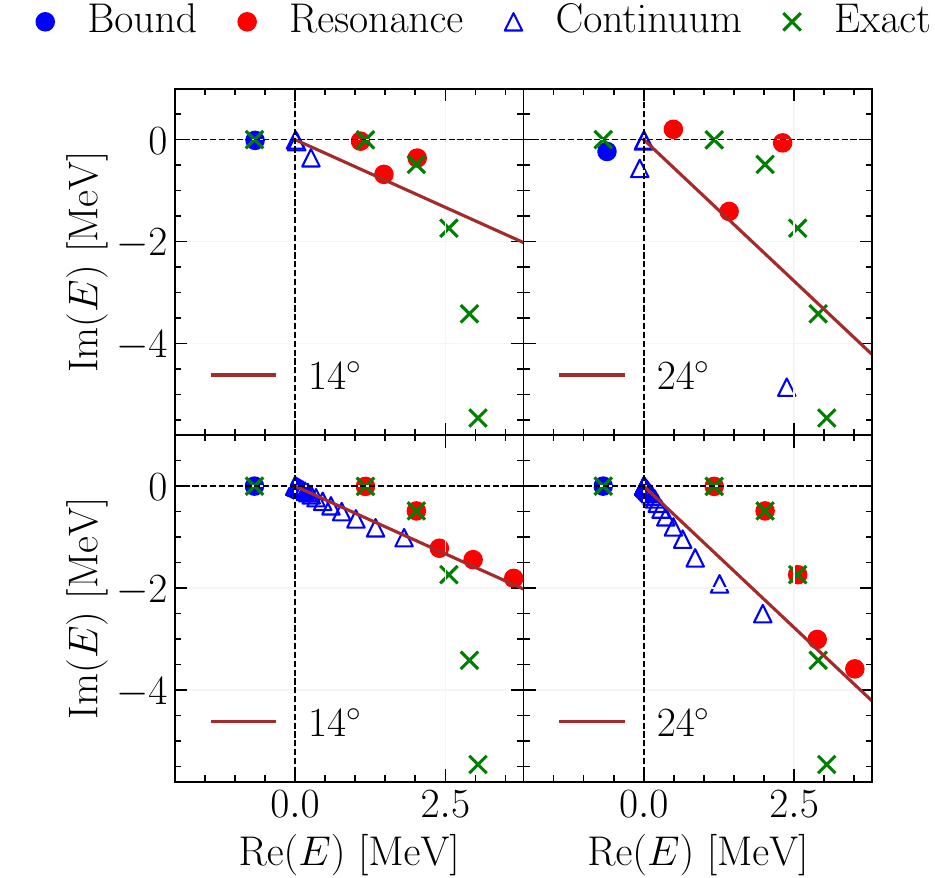}
    \caption{The spectra of the schematic Hamiltonian given by Eq.~\eqref{eq:schematic_potential_numerical} on the complex energy plane for the $1^{-}$ state. The top panel shows results for a basis size of $16$, and the bottom panel for a basis size of $75$. Green crosses, blue dots, brown solid lines, blue triangles and red points represent exact values from Ref.~\cite{mayo:1997ptp}, the bound states, the rotated continuum, the discrete continuum, and the calculated resonance positions, respectively.}
    \label{fig:cs_75bas_size_comb_24.00deg_ 1minus_state}
\end{figure}

\begin{figure}[htb!]
    \centering
    \includegraphics[width=\linewidth]{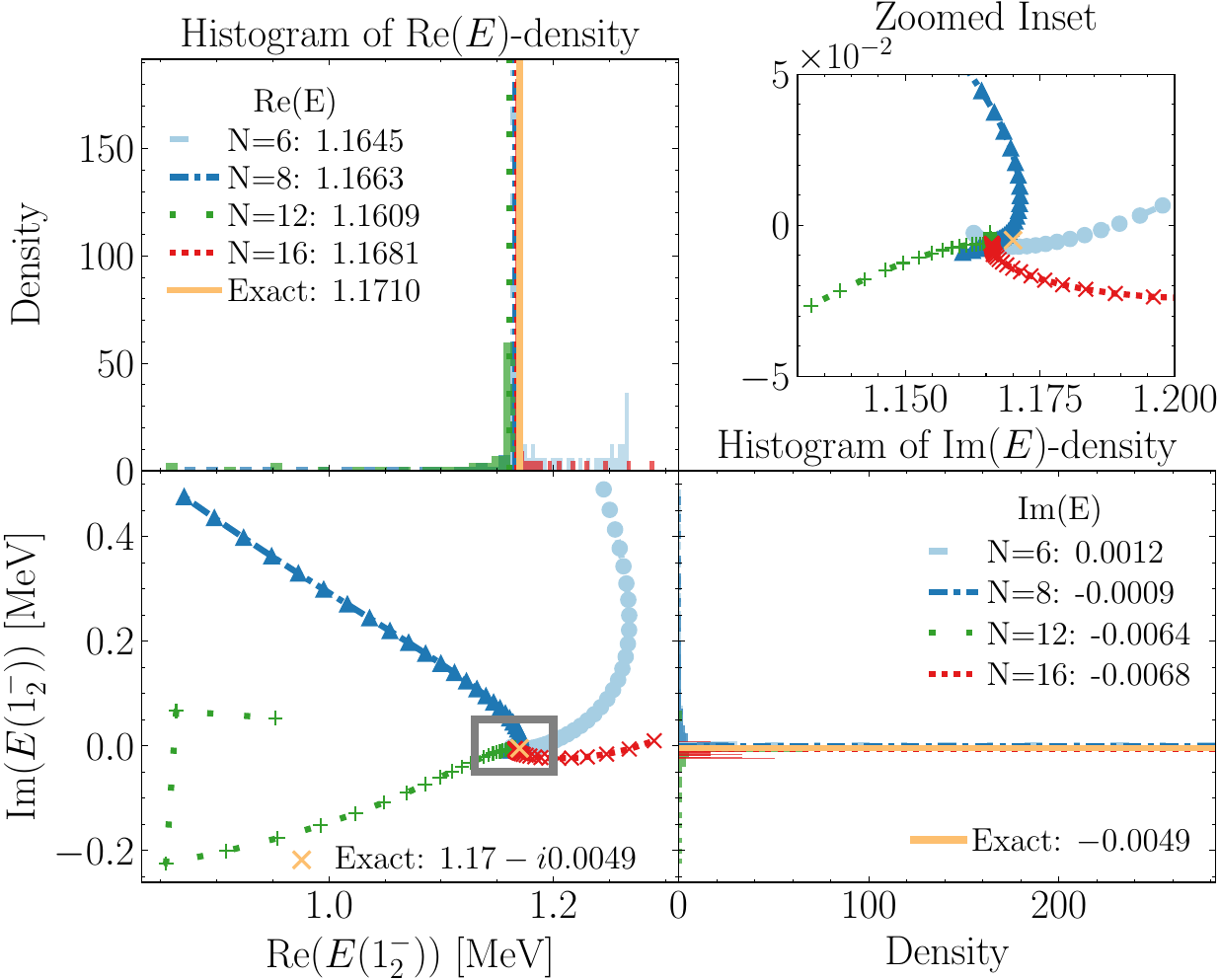}
    \caption{The $\theta$-trajectory for $N=6, 8, 12$, and $16$ for the $1^{-}_{2}$ state (i.e., the first resonance state) of the schematic potential given by Eq.~\eqref{eq:schematic_potential_numerical} is calculated using classical simulations. The value of $\theta$ ranges from $2^\circ$ to $45^\circ$ in steps of $0.5^\circ$. The density in the histogram represents the number of $\theta$-trajectory points falling within a particular bin. Vertical lines in the histogram indicate bins with the highest density of energy values, corresponding to the optimal resonance values. In the $\theta$-trajectory, the orange cross denotes the exact resonance position. Solid light orange lines in the histogram represent the exact real and imaginary components of the energy. The zoomed inset of a small rectangular region (shown in grey in $\theta$-trajectory) shows loops, kinks, or slowing down of the $\theta$-trajectory near the optimal resonance value.}
    \label{fig:theta_trej_N_6-16_E1-2}
\end{figure}
\begin{table}
\caption{Optimal resonance energies and absolute errors for the $1^{-}_{2}$ and $1^{-}_{3}$ states calculated using the $\theta$-trajectory method. Results are presented for various basis sizes ($N$). Energies are given in MeV, and exact value of $1^{-}_{2}$ and $1^{-}_{3}$ state are $1.1710 - i0.0049$ MeV and $2.0175 -i 0.4863$ MeV respectively, taken from Ref.~\cite{mayo:1997ptp}.}
\label{tab:table1}
\begin{ruledtabular}
\begin{tabular}{ccc} 
$N$ & $1^{-}_{2}$  & $1^{-}_{3}$ \\  \cline{2-3}
   & Energy [MeV]  & Energy [MeV] \\ \hline
$6$  & $1.1645 +i0.0012$ & $2.1048 -i0.6045$  \\ 
$8$  & $1.1661 -i0.0007$ & $2.0203 -i0.4822$  \\  
$12$ & $1.1601 -i0.0065$  & $2.0130 -i0.4609$  \\ 
$16$ & $1.1682 -i0.0067$  & $2.0120 -i0.4823$  \\ 
\end{tabular}
\end{ruledtabular}
\end{table}
As discussed in Sec.~\ref{subsec:optimal_basis}, the optimal set of Gaussian size parameters are those in geometric progression. For our calculation, the choice of parameters is $\qty{n_{\text{max}}, r_1, r_{n_{\text{max}}}}$, where $n_{\text{max}}$ represents the basis size.
We performed calculations for different basis sizes, ranging from $n_{\text{max}}=5$ to $n_{\text{max}}=75$, with $r_1 = 0.02$ fm and $r_{n_{\text{max}}}=n_{\text{max}}$ fm. We plot the shape of the schematic potential defined by Eq.~\eqref{eq:schematic_potential_numerical}, along with the bound state and the real parts of the first few resonance energies, in Fig.~\ref{fig:schemtaic_potential_complex}. The cyan-shaded region represents the discrete continuum arising from the finite basis size used in practical calculations. We calculate the resonances for the $0^{+}$ and $1^{-}$ states and focus only on the $1^{-}$ states for further calculations, as the first few resonances of the $1^{-}$ states lie farther from the origin compared to the $0^{+}$ states, making them easier to identify after rotating the continuum.
The complex energy eigenvalues are plotted in Fig.~\ref{fig:cs_75bas_size_comb_24.00deg_ 1minus_state} in the complex energy plane for two basis sizes, $N=16$ and $N=75$, with the two rotation angles (complex scaling parameter) $\theta$ of 14$^\circ$ and 24$^\circ$. The $N=75$ results demonstrate that all resonance positions are accurately reproduced and are almost independent of the $\theta$ value. In contrast, the upper panel of Fig.~\ref{fig:cs_75bas_size_comb_24.00deg_ 1minus_state} shows that a small basis size ($N=16$) is insufficient to reproduce resonance states accurately, and the resonance positions are highly dependent on $\theta$. Thus, as discussed in Sec.~\ref{sec:comp_res}, calculating the $\theta$-trajectory is crucial for determining the optimal resonance values of resonance position on complex energy plane.

We calculate the first resonance of the $1^-$ state, {\it i.e.} $1_{2}^{-}$, located around $1.2$ MeV. Fig.~\ref{fig:theta_trej_N_6-16_E1-2} shows the $\theta$-trajectory for this state, with basis sizes of $N=6,8,12$, and $16$. The value of $\theta$ ranges from $2^\circ$ to $45^\circ$ in steps of $0.5^\circ$.
These results show that even a small basis size of $N=6$ is sufficient to produce the first resonance state of $1^-$, with accuracy improving as the basis size increases. We observe bends and kinks near the optimal resonance position, which is clear from the zoomed inset of the optimal resonance region. However, for smaller basis sizes, such as $N=6$ or $8$, these features are not apparent at the optimal position for other resonance states, which is expected for such small basis sizes.

We provide an alternative method to extract the optimal resonance values in terms of histograms of the real and imaginary parts of the energy separately along the real and imaginary axes. The exact values of the real and imaginary energy components, obtained from Ref.~\cite{mayo:1997ptp} as a benchmark, are included in the histogram for reference. The highest density in the histogram corresponds to the optimal value of real and imaginary parts of complex energy eigenvalues.
With increasing basis size, the histogram peaks shift closer to the exact values and become sharper. This sharpening is evident in Fig.~\ref{fig:theta_trej_N_6-16_E1-2}, which is expected since the first resonance position requires a smaller basis size to achieve accurate results. These plots demonstrate that the optimal position of the first resonance of $1^{-}$ is accurately obtained, and the histogram further aids in determining the optimal position of the first resonance. Similarly, the optimal value of the second resonance of the $1^-$ state, i.e., $1^{-}_{3}$, is extracted using $\theta$-trajectory analysis for basis sizes $N=6$, $8$, $12$, and $16$. The compiled results are presented in Table~\ref{tab:table1}.


\subsubsection{Nucleus-nucleus scattering} \label{subsubsec:nucl_nucl_discu_classical}
\begin{figure}
    \centering
    \includegraphics[width=\linewidth]{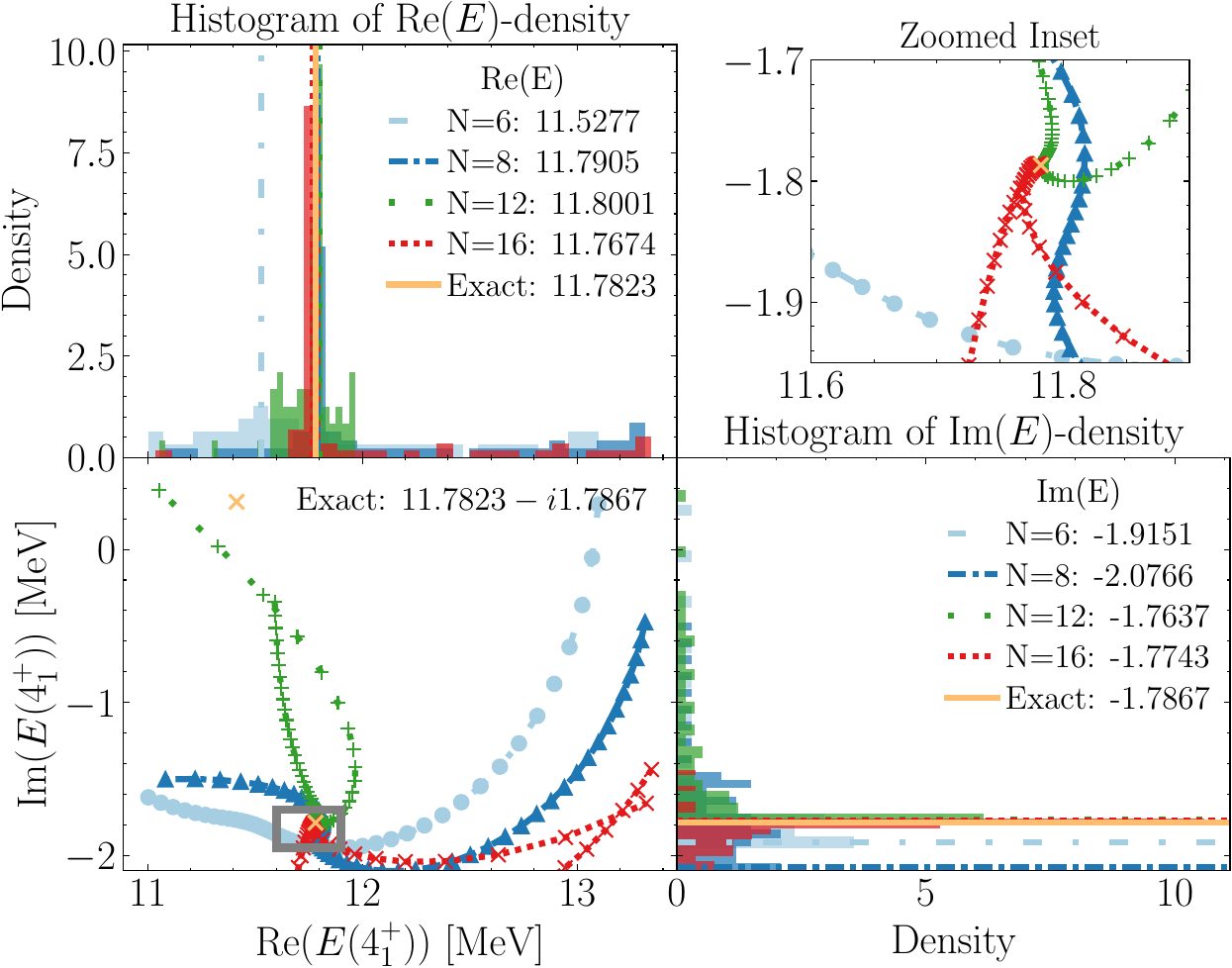}
    \caption{ Same as Fig.~\ref{fig:theta_trej_N_6-16_E1-2} but the $\theta$-trajectory for $N=6, 8, 12$, and $N=16$ for the $4^{+}_{1}$ state (i.e., the first resonance state) of $\alpha -\alpha$ scattering potential given by Eq.~\eqref{eq:nucleus-nucleus_pot} obtained using classical simulations.}
    \label{fig:theta_trej_classic_alp_alp_N_6-16_E4_plus}
\end{figure}
\begin{table*}[hbt!]
\caption{Optimal resonance energies and errors for the $0^{+}_{3}$, $2^{+}_{2}$, and $4^{+}_{3}$ states calculated using the $\theta$-trajectory method for $\alpha - \alpha$ model potential give by Eq.~\eqref{eq:nucleus-nucleus_pot}. Results are presented for various basis sizes ($N$). Energies are given in MeV, and exact values for $0^{+}_{3}$, $2^{+}_{2}$, and $4^{+}_{1}$ states are $0.092 -i0.29\times10^{-5}$  MeV, $2.88-i0.62$ MeV, and $11.78-i1.78$ MeV respectively.}
\label{tab:table2}
\begin{ruledtabular}
\begin{tabular}{ccccccc} 
\multirow{2}{*}{$N$} & $0^{+}_{3}$  & $2^{+}_{2}$ & $4^{+}_{1}$ \\  \cline{2-4}
   & Energy [MeV] & Energy [MeV] &  Energy [MeV]  \\ \hline
$6$  & $~~0.9168 +i0.4925$ &  $3.4824 -i0.0902$ & $11.5277 -i1.9151$ \\ 
$8$  & $~~0.4770 -i0.0093$ &  $3.1389 -i0.7618$ & $11.7905 -i2.0766$  \\  
$12$ & $-0.2863 -i0.0509$ & $2.9370 -i0.5348$ & $11.8001 -i1.7637$  \\ 
$16$ & $~~0.1126 -i0.0568$ &  $2.8766 -i0.5744$ & $11.7674 -i1.7743$  \\ 
$24$ & $~~0.1107 -i0.0157$ &  $2.8854 -i0.6098$ & $11.7932 -i1.7682$  \\
$32$ & $~~0.0848 -i0.0079$ &  $2.8907 -i0.6166$ & $11.7896 -i1.7655$  \\ 
\end{tabular}
\end{ruledtabular}
\end{table*}

As given in Sec.~\ref{subsec:nucleus_nucleus_pot}, we are using $\alpha-\alpha$ model potential given by Eq.~\eqref{eq:nucleus-nucleus_pot} to calculate $\alpha-\alpha$ two body resonance states. For our calculation, we use the HO basis given by Eq.~\eqref{eq:HO_basis}, with oscillator length parameter $b=1.36$ fm. 
This potential yields the three redundant bound states, two for $0^+$ and one for $2^+$ state. In addition to these redundant states, it has its first resonance state of $0^+$ at $0.092 -i0.29\times10^{-5}$  MeV, $2^+$ state at $2.88-i0.62$ MeV, while the first $4^+$ resonance state is at $11.78-i1.78$ MeV. Similar to Fig.~\ref{fig:schemtaic_potential_complex}, Fig.~\ref{fig:alp_alp_potential_complex_scaling} illustrates the $\alpha - \alpha$ model potential, displaying the bound state and real components of the first few resonance energies.
This section presents the results for $\theta$-trajectories of the first resonances of the $4^+$ state with basis sizes of $N=6, 8, 12$, and $16$. Fig.~\ref{fig:theta_trej_classic_alp_alp_N_6-16_E4_plus} shows the $\theta$-trajectory of the $4^{+}_{1}$ state, demonstrating that basis sizes $N=6$ and $8$ are insufficient to produce the optimal resonance position. As the basis size increases, the calculated resonance position approaches the exact one, which is evident from the shifting peaks of the histogram toward the exact result. With $N=16$, we can accurately reproduce the resonance position of the $4^{+}_{1}$ state. The zoomed inset of the $\theta$-trajectory near the exact resonance position clearly shows a kink at the optimal resonance position for $N=16$, further supported by our histogram for extracting the optimal resonance position.
Similarly, we calculate the $\theta$-trajectories of all three resonance states for basis sizes of $N=6,8,12,16,24$, and $32$, extract the optimal resonance positions using histograms, and present the results in Table~\ref{tab:table2}. For the $0^{+}_{3}$ state, $N=16$ is insufficient to obtain the optimal resonance position, while for $2^{+}_{2}$ and $4^{+}_{1}$ states, $N=16$ is large enough to achieve good agreement with true results. For the $0^{+}_{3}$ state, even with $N=32$, we do not obtain the exact decay width, although the real part of the energy is in good agreement. As the decay width of $0^{+}_{3}$ is extremely small, it is not accurately reproduced in comparison with experimental results.

\subsection{Nuclear resonances on a quantum computer}
In this section, we will discuss the result of both model potentials on a quantum simulator. From the discussion of the previous section, we know about the minimum basis size required to obtain optimal resonance position. With known basis size, we simulate the results on quantum simulator using Algorithm $1$ and Algorithm $2$.

\subsubsection{Schematic potential}   \label{subsec:schematic_pot_quant}
\begin{figure}
    \centering
    \includegraphics[width=\linewidth]{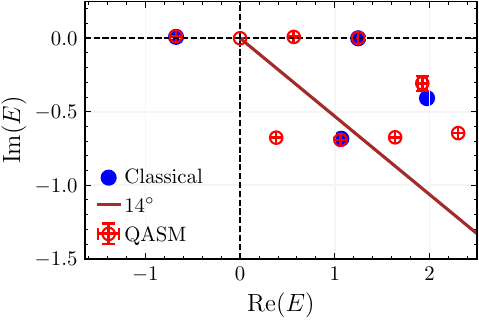}
    \caption{The complex energy spectrum for the schematic potential on complex energy plane, as defined by Eq.~\eqref{eq:schematic_potential_numerical}, is shown for a basis size of $N=5$ with Jordan-Wigner (JW) transformation. The blue circles represent the classical eigenvalues, and the red plus symbols indicate the eigenvalues calculated on a QASM simulator using VQA. Energies are median of $120$ independent runs, with error as MAD. These calculations employ the \emph{Ans\"atze} described by Eq.~\eqref{eq:ansatz}, with the BFGS optimizer used for the variational optimization.}
    \label{fig:Energy_complex_plain_N5_L3_P3_JW_optimizer_BFGS_qasm_xu_xie_median_outliner}
\end{figure}

In Sec.~\ref{subsec:var_alg_quant}, Algorithm $1$ is devised to calculate the complex energy spectrum on a quantum computer. We use this algorithm to compute the energy spectrum of the complex-rotated schematic Hamiltonian, and plot the results in Fig.~\ref{fig:Energy_complex_plain_N5_L3_P3_JW_optimizer_BFGS_qasm_xu_xie_median_outliner} for $N=5$ using Gaussian basis and JW mapping. From Sec.~\ref{subsubsec:scematic_discu_classical}, it is clear that $N=6$ is large enough to obtain the resonance position. To reduce the number of qubits further, we performed calculations with $N=5$ on a quantum simulator through which we are still able to obtain the resonance position. Fig.~\ref{fig:Energy_complex_plain_N5_L3_P3_JW_optimizer_BFGS_qasm_xu_xie_median_outliner} demonstrates that the complete spectrum can be reproduced by running Algorithm $1$ with an appropriate initial guess and step size. We start with an initial guess of $E_r = -1.0$ MeV and $E_i = -0.01$ MeV, repeating the algorithm $40$ times with a step size of $0.4$ MeV for the real part of the energy ($E_r$). A small step size is chosen to ensure the complete generation of the spectrum. We use the \emph{Ans\"atze} given by Eq.~\eqref{eq:ansatz} with $P=3$, which consists of three layers of the \emph{Ans\"atze} block defined in Eq.~\eqref{eq:ansatz_single_block}. The choice of $P=3$ is motivated by the fact that increasing the number of layers allows the \emph{Ans\"atze} to explore a larger Hilbert space, thereby improving the accuracy of the results. However, this increased accuracy comes at the cost of a greater circuit depth and a larger number of parameters to optimize. For $P < 3$, the algorithm fails to converge to the desired accuracy, making $P=3$ the optimal choice for our calculations. 
For this calculation, we use the IBM Qiskit software package~\cite{qiskit2024}, which supports several backends, such as the statevector simulator and the QASM (Quantum Assembly Language) simulator. The QASM simulator mimics an ideal quantum computer for quantum simulations. We use the QASM simulator for our quantum simulation, executing $120$ independent runs of VQA with 8192 shots of the simulator for each run. Our calculation employs the \emph{Ans\"atze} described by Eq.~\eqref{eq:ansatz}, with the Broyden-Fletcher-Goldfarb-Shanno (BFGS) optimizer used for the variational optimization. We choose the BFGS optimizer for our calculations, as it yields the best results among the optimizers available in the Python SciPy library~\cite{2020SciPy-NMeth}. We use the median value of $120$ runs instead of the mean, as one outlier in the data could skew the estimation of the result. Correspondingly, we use the median absolute deviation (MAD) for error calculation.

 From Fig.~\ref{fig:Energy_complex_plain_N5_L3_P3_JW_optimizer_BFGS_qasm_xu_xie_median_outliner}, it is observed that, in addition to the five existing eigenvalues obtained by diagonalizing the Hamiltonian matrix, there are additional ones obtained by quantum simulation. These extra eigenvalues are obtained after applying the Jordan-Wigner (JW) transformation to the Hamiltonian. However, the JW transformation does not conserve particle number, which is a conserved quantity in our system. Since the \emph{Ans\"atze} used in our calculation searches for eigenstates in a larger Hilbert space, it leads to the inclusion of these extra states as eigenstates of the Hamiltonian. To address this issue, three potential solutions can be considered:
\begin{enumerate}
    \item We can choose an appropriate \emph{Ans\"atze} that preserves particle number conservation. 
    \item We can project the calculated eigenstate onto a particle-number-conserving state and filter out states with larger fidelity. This method is discussed in detail in Sec.~\ref{subsec:state_fileteration}.
    \item Alternatively, we can transform the Hamiltonian using a more efficient transformation, such as the GC  transformation, which reduces the number of qubits exponentially and automatically filters out the extra states.
\end{enumerate}

\begin{figure}
    \centering
    \includegraphics[width=\linewidth]{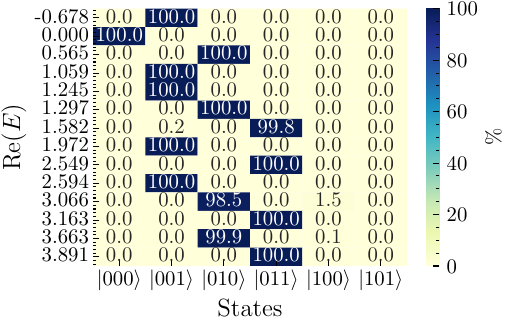}
    \caption{The heatmap represents the percentage distribution of measured ancilla states corresponding to single-particle projections for each energy level. The real part of the energy is shown on the y-axis, while the x-axis denotes the states. Each cell indicates the relative count of a specific state, normalized by the total counts for the corresponding energy level.}
    \label{fig:redundant_state_sorting_jw_N5_n40_de0.4_step0.2_igr_minus1.0_percent_final}
\end{figure}

Here we explore the last two methods. We start by projecting the eigenstate obtained from VQA calculations onto a single-particle state employing the quantum phase estimation (QPE) algorithm using the unitary operator $U_N$ derived from the number operator as discussed in Sec~\ref{subsec:state_fileteration}. After applying VQA, the resulting wave function is subjected to QPE with operator $U_N$ and $n_r$ number of ancilla qubits. The final state becomes $\sum_i c_i\ket{\theta_i 2^{n_r}}\otimes\ket{\phi_i}$, where $\theta_i 2^{n_r}$ represents the particle number information encoded in binary. In our case, for a single-particle system distributed across $N$ basis states, the state $\ket{001}\otimes\ket{\phi_i}$ corresponds to the particle-number-conserving eigenstate, while other states are considered redundant. In calculations, we feed each eigenstate obtained after VQA to QPE as the initial state and measure the ancilla qubits for each energy, and extract the counts. Finally, we plot the percentage of counts for different energy in different states, corresponding to ancilla qubit, as a heatmap in Fig.~\ref{fig:redundant_state_sorting_jw_N5_n40_de0.4_step0.2_igr_minus1.0_percent_final}. It illustrates that only for energy eigenvalues corresponding to classical eigenvalues, the state $\ket{001}$ exhibits a $100\%$ count. For other energy eigenvalues, the counts are predominantly distributed among other states, with minimal presence in $\ket{001}$. This observation leads to the conclusion that only these five states are particle-conserving, while the others are identified as redundant states.      
This projection method effectively filters out states that do not preserve the single-particle criterion, ensuring that only physically relevant states are retained for further analysis. Our results emphasize the significance of utilizing system symmetries in quantum simulations and provide a reliable method for improving the accuracy and efficiency of excited-state calculations in fermionic systems. 

\begin{figure}
    \centering
    \includegraphics[width=\linewidth]{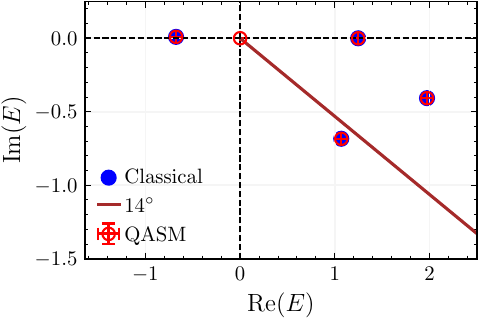}
    \caption{Same as Fig.~\ref{fig:Energy_complex_plain_N5_L3_P3_JW_optimizer_BFGS_qasm_xu_xie_median_outliner} but with GC encoding.}
    \label{fig:Energy_complex_plain_N5_L3_P3_GC_optimizer_BFGS_qasm_xu_xie_median_outliner}
\end{figure}

In addition, GC encoding and its corresponding transformation, as suggested in point three, can be utilized. The GC transformation requires fewer qubits and, when applied carefully, prevents the introduction of redundant states. For instance, for $N=8$, only three qubits are needed for mapping the basis states to qubits, ensuring that the total number of states remains eight after the transformation. In the present case of $N=5$, GC encoding also requires only three qubits, and any additional states appear at the origin, effectively eliminating them.
We plot the spectrum of the complex-rotated Hamiltonian after applying the GC transformation in Fig.~\ref{fig:Energy_complex_plain_N5_L3_P3_GC_optimizer_BFGS_qasm_xu_xie_median_outliner}. It is evident that no extra states are introduced, apart from one at the origin, which can be discarded. To obtain the full spectrum, we employ Algorithm $1$ and repeat the steps $20$ times with a step size of $0.4$ MeV. The results simulated on a QASM simulator exhibit excellent agreement with the classically diagonalized results.
Thus, GC encoding, being more efficient, is our preferred choice for further calculations, as it significantly reduces the number of qubits required for simulation. Using this approach, we can perform calculations with a basis size of $N=32$ using only five qubits, whereas the JW transformation would require $32$ qubits, which increases the complexity of the simulation due to a larger \emph{Ans\"atze} and an excessive number of parameters.

\begin{figure}[hbt!]
    \centering
    \includegraphics[width=\linewidth]{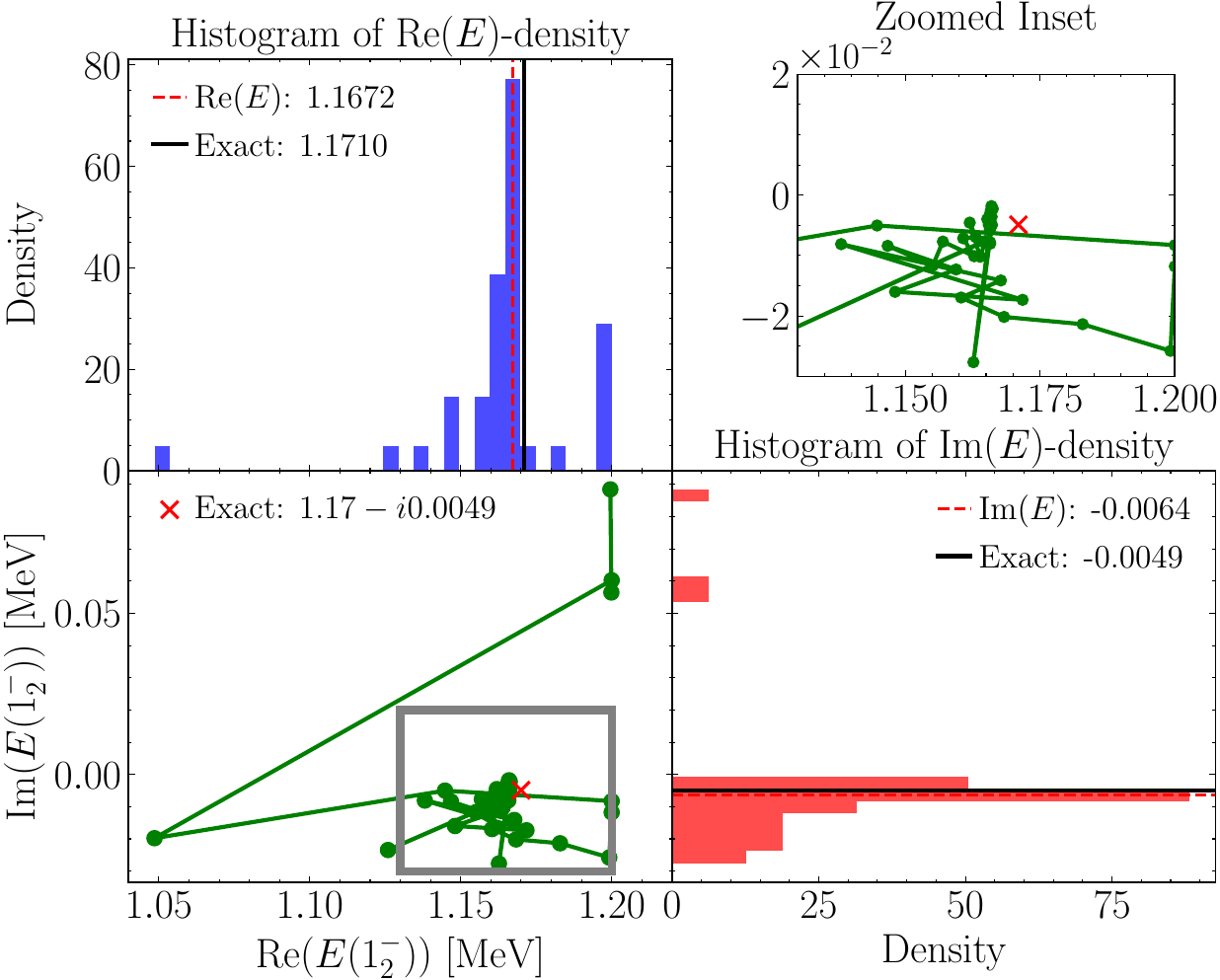}
    \caption{The $\theta$-trajectory for $N=16$ for the $1^{-}_{2}$ state is shown. The value of $\theta$ is taken from $\theta=2^\circ$ to $30^\circ$ in a step size of $0.5^\circ$ each. The vertical lines in the histogram represent the bins with the highest density of energy values, corresponding to the optimal resonance values. In the $\theta$-trajectory, the red cross indicates the exact resonance position. Additionally, black solid lines are drawn in the histogram to represent the exact real and imaginary components of the energy. This calculation uses GC transformation and employs the \emph{Ans\"atze} described by Eq.~\eqref{eq:ansatz}, with the BFGS optimizer used for the variational optimization.}
    \label{fig:theta_trej_quant_N_16_L4_P3_E1-2}
\end{figure}

Having successfully generated the complete spectrum of the complex-rotated Hamiltonian on a quantum simulator, we proceed to compute the $\theta$-trajectory to determine the optimal resonance position purely through quantum simulation. To generate the $\theta$-trajectory, we employ Algorithm $2$ as described in Sec.~\ref{subsec:theta_traj_quant}.
From Fig.~\ref{fig:Energy_complex_plain_N5_L3_P3_GC_optimizer_BFGS_qasm_xu_xie_median_outliner}, it is clear that the first resonance lies near $1.1$ MeV, with a very small imaginary part indicative of a narrow decay width. Accordingly, we set the initial guess for $E_r$ and $E_i$ as $1.1$ MeV and $0.0$ MeV, respectively, and execute Algorithm $1$ for each specified value of $\theta$. For each $\theta$, we select results from Algorithm $1$ that fall within a small neighborhood of $1.1 + i0.0$ MeV. This process is repeated for $\theta$ values ranging from $0^\circ$ to $30^\circ$, in increments of $0.5^\circ$.
The $\theta$-trajectory results for the $1_{2}^{-}$ state, obtained using the quantum simulator, are shown in Fig.~\ref{fig:theta_trej_quant_N_16_L4_P3_E1-2}. This result is obtained using GC encoding and corresponding transformation, and four qubits are used for quantum simulations using statevector simulator. One can also observe that the number of point is lower than expected for the $\theta$-trajectory. This occurs because for certain values of $\theta$, Algorithm $1$ converges to an energy value that does not lie within the small neighborhood of $1.1 + i0.0$ MeV. As a result, a well-defined $\theta$-trajectory is not obtained in this case. However, the optimal resonance position can still be extracted with the aid of a histogram. These results demonstrate that the optimal resonance energy is accurately determined and is in good agreement with the exact value. Slight deviations of some resonance positions from the actual values can also be observed because we are using an optimization routine to minimize the variance of Hamiltonian. This could lead to a significantly small variation in the final result as the energy is calculated as a parameter.
Similar to Fig.~\ref{fig:theta_trej_quant_N_16_L4_P3_E1-2}, we compute the results for the $1_{3}^{-}$ state using a basis size of $N=16$. In this case, a well-defined $\theta$-trajectory is obtained, and the optimal resonance position is extracted and presented in Table~\ref{tab:table3_quant}. Thus, the optimal positions of the first two $1^{-}$ resonance states have been successfully extracted, which are in good agreement with the exact values reported in Ref.~\cite{mayo:1997ptp}.
Moreover, by utilizing the histogram analysis, we successfully extracted the optimal resonance positions, which are in close agreement with the true resonance values.
\begin{table}[hbt!]
\caption{Optimal resonance energies and absolute errors for the $1^{-}_{2}$ and $1^{-}_{3}$ states calculated using the $\theta$-trajectory method on quantum simulator for basis size $N=16$. Energies are given in MeV, and errors are calculated with respect to true values given in Table~\ref{tab:table1}, taken from Ref.~\cite{mayo:1997ptp}. The error is defined as the absolute modulus of the complex difference between the obtained and true energy eigenvalues.}
\label{tab:table3_quant}
\begin{ruledtabular}
\begin{tabular}{cccc} 
\multirow{2}{*}{State} & \multicolumn{2}{c}{Energy [MeV]} & \multirow{2}{*}{Error} \\  \cline{2-3}
   & True value & Quantum simulator  &  \\ \hline
$1^{-}_{2}$  & 1.1710 -$i$0.0049 & 1.1672 -$i$0.0064 & 0.0032  \\ 
$1^{-}_{3}$  & 2.0175 -$i$0.4863 & 2.0065 -$i$0.4732 & 0.0215 \\  
\end{tabular}
\end{ruledtabular}
\end{table}

In the next section, we implement the same methodology to a much more realistic potential to solve for the resonance states of $\alpha - \alpha$ scattering. 

\subsubsection{Nucleus-nucleus scattering}
\begin{figure}
    \centering
    \includegraphics[width=\linewidth]{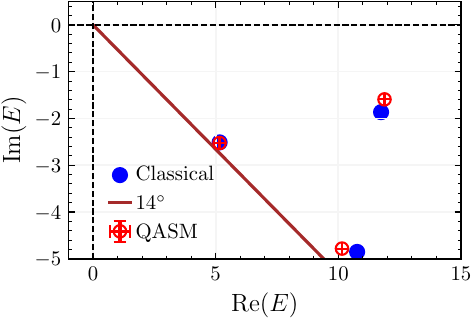}
    \caption{Same as Fig.~\ref{fig:Energy_complex_plain_N5_L3_P3_JW_optimizer_BFGS_qasm_xu_xie_median_outliner} but for $4^{+}_{1}$ state of the nucleus-nucleus scattering potential defined by Eq.~\eqref{eq:nucleus-nucleus_pot} using GC encoding and a basis size of $N=16$.}
    \label{fig:Energy_complex_plain_N16_L4_P3_GC_4plus_optimizer_BFGS_qasm_xu_xie_median}
\end{figure}
This section presents the result of quantum simulations on the nucleus-nucleus potential to extract optimal resonance position. As discussed in Sec.~\ref{subsec:schematic_pot_quant}, we utilize Algorithm $1$ to compute the energy spectrum of the complex-rotated nucleus-nucleus scattering Hamiltonian using the GC transformation. The results are plotted in Fig.~\ref{fig:Energy_complex_plain_N16_L4_P3_GC_4plus_optimizer_BFGS_qasm_xu_xie_median} for $4^{+}_{1}$ state with $N=16$ basis size using the HO basis. The complete spectrum is generated by running Algorithm $1$ with an appropriate initial guess and step size. We initiate the calculation with an initial guess of $E_r = 5.0$ MeV and $E_i = -0.01$ MeV, repeating the algorithm $20$ times with a step size of $1.0$ MeV for the real part of the energy ($E_r$). Although a small step size is typically chosen for precise spectrum generation, in this case, a step size of $1.0$ MeV is used because with HO basis energy is equispaced and in our case, even continuum will be widely spaced and is not densely populated at the origin, which is the case for Gaussian basis~\cite{Moiseyev:mole1978}.
From Fig.~\ref{fig:Energy_complex_plain_N16_L4_P3_GC_4plus_optimizer_BFGS_qasm_xu_xie_median}, it is clear that the first resonance of $4^{+}$ state lies around $11.88 -i1.59$ MeV, in complex energy plane. Thus to extract the optimum value of resonance, we compute the $\theta$-trajectory of $4_{1}^{+}$ state. 
\begin{figure}[hbt!]
    \centering
    \includegraphics[width=\linewidth]{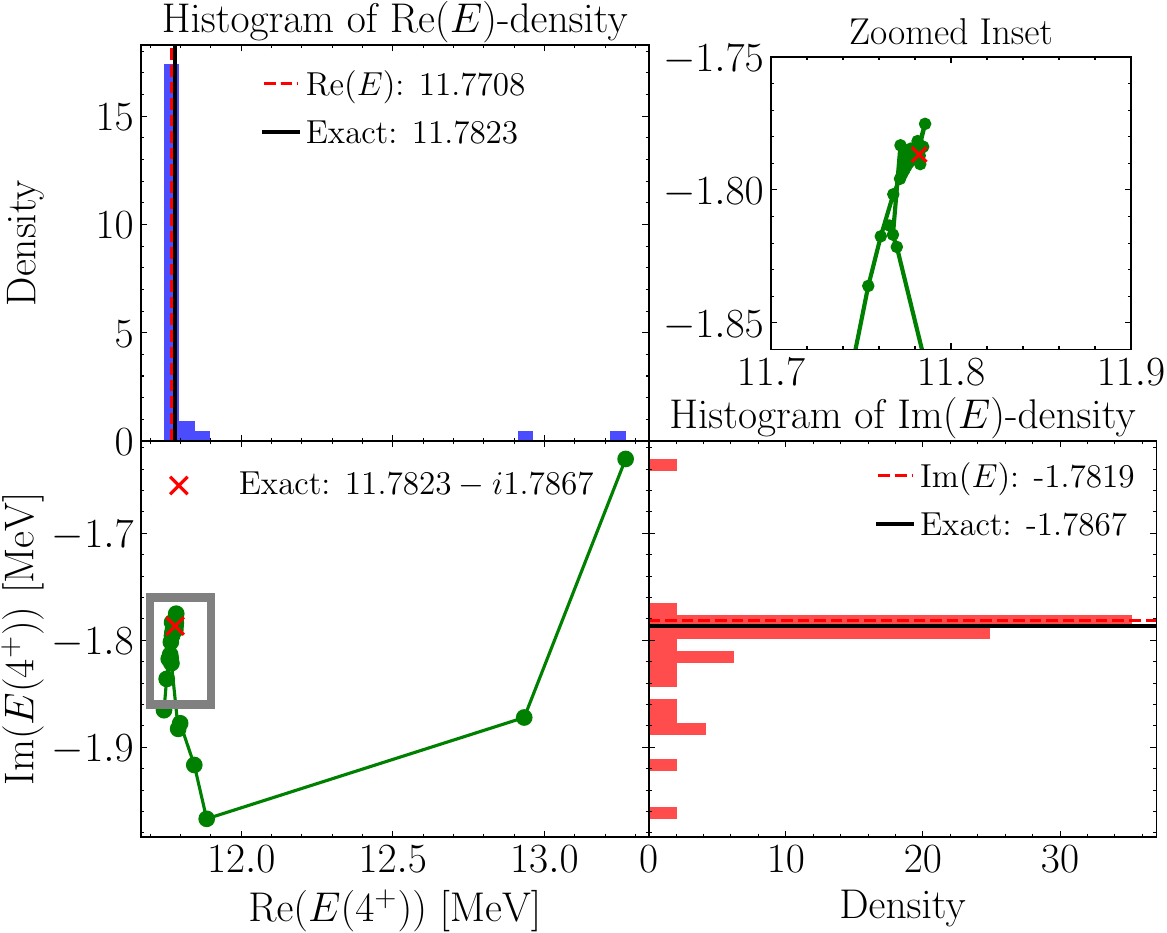}
    \caption{Same as Fig.~\ref{fig:theta_trej_quant_N_16_L4_P3_E1-2} but for the $4^{+}_{1}$ state of nucleus-nucleus scattering potential.}
    \label{fig:theta_trej_quant_alp_alp_N_16_L4_P3_E_4_plus_GC_statevec}
\end{figure}

Similar to Sec.~\ref{subsec:schematic_pot_quant}, we have used Algorithm $2$ to generate the $\theta$-trajectory for $4^{+}_{1}$ state and plotted in Fig.~\ref{fig:theta_trej_quant_alp_alp_N_16_L4_P3_E_4_plus_GC_statevec}. To obtain proper $\theta$-trajectory, we use \emph{Ans\"atze} with $P=4$ to achieve the desired accuracy. We start with an initial guess of $11.5$ MeV for $E_r$ and $-0.01$ MeV for $E_i$ with basis size $16$ using GC transformation and execute Algorithm $1$ for each specified value of $\theta$. For each $\theta$, we select results that fall within a small neighborhood of $1.1 + i0.0$ MeV. This process is repeated for $\theta$ values ranging from $0^\circ$ to $30^\circ$, in increments of $0.5^\circ$. A well-defined $\theta$-trajectory is obtained, although with fewer points than expected. Similar to the discussion in Sec.~\ref{subsec:schematic_pot_quant}, we extract only those resonance positions that are in the neighborhood of $1.1 + i0.0$ MeV, while a few points converge to other values for a given $\theta$. Additionally, smaller values of $\theta$ are insufficient to extract this resonance position, which reduces the number of points. From Fig.~\ref{fig:theta_trej_quant_alp_alp_N_16_L4_P3_E_4_plus_GC_statevec}, it is evident that the extracted resonance position, aided by a histogram, is in good agreement with the exact value obtained classically using a larger basis size. Following the same procedure, we have extracted the optimal resonance position for the $2^{+}_{2}$ state and present the results for both resonance positions in Table~\ref{tab:table4_quant}. From Table~\ref{tab:table4_quant}, it is clear that we are able to extract the optimal value of resonance position on complex energy plane purely through a quantum simulator.

\begin{table}[hbt!]
\caption{Similar to Table.~\ref{tab:table3_quant}, we present optimal resonance energies and absolute errors for the $2^{+}_{2}$ and $4^{+}_{1}$ states calculated using the $\theta$-trajectory method on quantum simulator for basis size $N=16$. Errors are calculated with respect to true values calculated classically with larger basis size as mentioned in Table~\ref{tab:table2}.}
\label{tab:table4_quant}
\begin{ruledtabular}
\begin{tabular}{cccc} 
\multirow{2}{*}{State} & \multicolumn{2}{c}{Energy [MeV]} & \multirow{2}{*}{Error} \\  \cline{2-3}
   & True value & Quantum simulator  &  \\ \hline
$2^{+}_{2}$  & $2.8932 -i0.6224$ & $2.6913 - i0.5816$ &  $0.2059$ \\ 
$4^{+}_{1}$  & $11.7823 -i1.7867$ & $11.7840 -i1.7639$ & $0.0228$ \\  
\end{tabular}
\end{ruledtabular}
\end{table}

\section{Conclusions} \label{sec:conclussion}
We present the first quantum simulation of nuclear resonances using variational algorithms in conjunction with a resource-efficient GC encoding. By leveraging noise-resilient protocols, we demonstrate that NISQ-era quantum devices can effectively simulate non-Hermitian Hamiltonians with a remarkable degree of accuracy, paving the way for quantum computation of complex nuclear phenomena.

To simulate nuclear resonances on a quantum computer, we begin by applying a complex rotation to the Hamiltonian, rendering it non-Hermitian. Using an appropriately chosen basis, we first obtain the $\theta$-trajectory classically for various basis sizes $(N)$ corresponding to the resonance states of a schematic potential. We then extend this approach to simulate the resonance states of a more realistic $\alpha - \alpha$ scattering potential to determine the optimal resonance position. This process allows us to identify the minimum basis size required for accurately extracting the resonance position. For these smaller basis sizes amenable to quantum computation, we deploy Algorithm $1$ to compute the complex energy spectrum and isolate resonance positions on a quantum simulator. Algorithm $2$ is then used to construct the $\theta$-trajectories and extract optimal resonance energies. As a proof-of-concept, we compute the D- and G-wave resonances of $\alpha-\alpha$ scattering potential, achieving excellent agreement with classical benchmark results.

This study represents a crucial first step toward computing nuclear resonances entirely on a quantum computer by extracting optimal resonance positions using the $\theta$-trajectory method. Beyond nuclear physics, the protocols introduced here offer a versatile and scalable framework for simulating resonance phenomena in atomic, molecular, and condensed matter systems. Several promising directions for future exploration remain. One particularly interesting avenue is to bypass extensive $\theta$-trajectory calculations by exploring direct search methods, which could offer improved convergence properties over the current approach. Our results lay the groundwork for quantum-enabled studies of open quantum systems, and mark a significant advancement in bridging quantum information science with nuclear theory.

\begin{acknowledgements}
This work is supported by the SERB-DST, Govt.~of India, via project \sloppy{CRG/2022/009359}.
We acknowledge the National Supercomputing Mission (NSM) for providing computing resources of `PARAM Ganga' at IIT Roorkee, which is implemented by C-DAC and supported by MeitY and DST, Govt.~of India. This work was performed under the auspices of the U.S. Department of Energy by Lawrence Livermore National Laboratory under Contract No. DE-AC52-07NA27344. We thank Aman Gupta for early discussions on the direct measurement approach to solving complex energy eigenvalue problems.
\end{acknowledgements}

\bibliographystyle{apsrev}
\bibliography{reference}

\providecommand{\noopsort}[1]{}\providecommand{\singleletter}[1]{#1}%
\begin{thebibliography}{44}
\expandafter\ifx\csname natexlab\endcsname\relax\def\natexlab#1{#1}\fi
\expandafter\ifx\csname bibnamefont\endcsname\relax
  \def\bibnamefont#1{#1}\fi
\expandafter\ifx\csname bibfnamefont\endcsname\relax
  \def\bibfnamefont#1{#1}\fi
\expandafter\ifx\csname citenamefont\endcsname\relax
  \def\citenamefont#1{#1}\fi
\expandafter\ifx\csname url\endcsname\relax
  \def\url#1{\texttt{#1}}\fi
\expandafter\ifx\csname urlprefix\endcsname\relax\def\urlprefix{URL }\fi
\providecommand{\bibinfo}[2]{#2}
\providecommand{\eprint}[2][]{\url{#2}}

\bibitem[{\citenamefont{Bharti et~al.}(2022)\citenamefont{Bharti,
  Cervera-Lierta, Kyaw, Haug, Alperin-Lea, Anand, Degroote, Heimonen, Kottmann,
  Menke et~al.}}]{RevModPhys.94.015004_9}
\bibinfo{author}{\bibfnamefont{K.}~\bibnamefont{Bharti}},
  \bibinfo{author}{\bibfnamefont{A.}~\bibnamefont{Cervera-Lierta}},
  \bibinfo{author}{\bibfnamefont{T.~H.} \bibnamefont{Kyaw}},
  \bibinfo{author}{\bibfnamefont{T.}~\bibnamefont{Haug}},
  \bibinfo{author}{\bibfnamefont{S.}~\bibnamefont{Alperin-Lea}},
  \bibinfo{author}{\bibfnamefont{A.}~\bibnamefont{Anand}},
  \bibinfo{author}{\bibfnamefont{M.}~\bibnamefont{Degroote}},
  \bibinfo{author}{\bibfnamefont{H.}~\bibnamefont{Heimonen}},
  \bibinfo{author}{\bibfnamefont{J.~S.} \bibnamefont{Kottmann}},
  \bibinfo{author}{\bibfnamefont{T.}~\bibnamefont{Menke}},
  \bibnamefont{et~al.}, \bibinfo{journal}{Rev. Mod. Phys.}
  \textbf{\bibinfo{volume}{94}}, \bibinfo{pages}{015004}
  (\bibinfo{year}{2022}),
  \urlprefix\url{https://link.aps.org/doi/10.1103/RevModPhys.94.015004}.

\bibitem[{\citenamefont{Savage}(2024)}]{refId0_martin_18}
\bibinfo{author}{\bibfnamefont{M.~J.} \bibnamefont{Savage}},
  \bibinfo{journal}{EPJ Web Conf.} \textbf{\bibinfo{volume}{296}},
  \bibinfo{pages}{01025} (\bibinfo{year}{2024}),
  \urlprefix\url{https://doi.org/10.1051/epjconf/202429601025}.

\bibitem[{\citenamefont{McClean et~al.}(2016)\citenamefont{McClean, Romero,
  Babbush, and Aspuru-Guzik}}]{McClean_2016_19}
\bibinfo{author}{\bibfnamefont{J.~R.} \bibnamefont{McClean}},
  \bibinfo{author}{\bibfnamefont{J.}~\bibnamefont{Romero}},
  \bibinfo{author}{\bibfnamefont{R.}~\bibnamefont{Babbush}}, \bibnamefont{and}
  \bibinfo{author}{\bibfnamefont{A.}~\bibnamefont{Aspuru-Guzik}},
  \bibinfo{journal}{New Journal of Physics} \textbf{\bibinfo{volume}{18}},
  \bibinfo{pages}{023023} (\bibinfo{year}{2016}),
  \urlprefix\url{https://dx.doi.org/10.1088/1367-2630/18/2/023023}.

\bibitem[{\citenamefont{Baroni et~al.}(2022)\citenamefont{Baroni, Carlson,
  Gupta, Li, Perdue, and Roggero}}]{PhysRevD.105.074503_4}
\bibinfo{author}{\bibfnamefont{A.}~\bibnamefont{Baroni}},
  \bibinfo{author}{\bibfnamefont{J.}~\bibnamefont{Carlson}},
  \bibinfo{author}{\bibfnamefont{R.}~\bibnamefont{Gupta}},
  \bibinfo{author}{\bibfnamefont{A.~C.~Y.} \bibnamefont{Li}},
  \bibinfo{author}{\bibfnamefont{G.~N.} \bibnamefont{Perdue}},
  \bibnamefont{and} \bibinfo{author}{\bibfnamefont{A.}~\bibnamefont{Roggero}},
  \bibinfo{journal}{Phys. Rev. D} \textbf{\bibinfo{volume}{105}},
  \bibinfo{pages}{074503} (\bibinfo{year}{2022}),
  \urlprefix\url{https://link.aps.org/doi/10.1103/PhysRevD.105.074503}.

\bibitem[{\citenamefont{Turro et~al.}(2024)\citenamefont{Turro, Wendt,
  Quaglioni, Pederiva, and
  Roggero}}]{turro2024evaluationphaseshiftsnonrelativistic_6}
\bibinfo{author}{\bibfnamefont{F.}~\bibnamefont{Turro}},
  \bibinfo{author}{\bibfnamefont{K.~A.} \bibnamefont{Wendt}},
  \bibinfo{author}{\bibfnamefont{S.}~\bibnamefont{Quaglioni}},
  \bibinfo{author}{\bibfnamefont{F.}~\bibnamefont{Pederiva}}, \bibnamefont{and}
  \bibinfo{author}{\bibfnamefont{A.}~\bibnamefont{Roggero}},
  \emph{\bibinfo{title}{Evaluation of phase shifts for non-relativistic elastic
  scattering using quantum computers}} (\bibinfo{year}{2024}),
  \eprint{2407.04155}, \urlprefix\url{https://arxiv.org/abs/2407.04155}.

\bibitem[{\citenamefont{Du et~al.}(2021)\citenamefont{Du, Vary, Zhao, and
  Zuo}}]{PhysRevA.104.012611_scattering_7}
\bibinfo{author}{\bibfnamefont{W.}~\bibnamefont{Du}},
  \bibinfo{author}{\bibfnamefont{J.~P.} \bibnamefont{Vary}},
  \bibinfo{author}{\bibfnamefont{X.}~\bibnamefont{Zhao}}, \bibnamefont{and}
  \bibinfo{author}{\bibfnamefont{W.}~\bibnamefont{Zuo}},
  \bibinfo{journal}{Phys. Rev. A} \textbf{\bibinfo{volume}{104}},
  \bibinfo{pages}{012611} (\bibinfo{year}{2021}),
  \urlprefix\url{https://link.aps.org/doi/10.1103/PhysRevA.104.012611}.

\bibitem[{\citenamefont{Peruzzo et~al.}(2014)\citenamefont{Peruzzo, McClean,
  Shadbolt, Yung, Zhou, Love, Aspuru-Guzik, and O’Brien}}]{nature_23}
\bibinfo{author}{\bibfnamefont{A.}~\bibnamefont{Peruzzo}},
  \bibinfo{author}{\bibfnamefont{J.}~\bibnamefont{McClean}},
  \bibinfo{author}{\bibfnamefont{P.}~\bibnamefont{Shadbolt}},
  \bibinfo{author}{\bibfnamefont{M.-H.} \bibnamefont{Yung}},
  \bibinfo{author}{\bibfnamefont{X.-Q.} \bibnamefont{Zhou}},
  \bibinfo{author}{\bibfnamefont{P.~J.} \bibnamefont{Love}},
  \bibinfo{author}{\bibfnamefont{A.}~\bibnamefont{Aspuru-Guzik}},
  \bibnamefont{and} \bibinfo{author}{\bibfnamefont{J.~L.}
  \bibnamefont{O’Brien}}, \bibinfo{journal}{Nature Communications}
  \textbf{\bibinfo{volume}{5}}, \bibinfo{pages}{2041} (\bibinfo{year}{2014}),
  \urlprefix\url{https://doi.org/10.1038/ncomms5213}.

\bibitem[{\citenamefont{Dumitrescu et~al.}(2018)\citenamefont{Dumitrescu,
  McCaskey, Hagen, Jansen, Morris, Papenbrock, Pooser, Dean, and
  Lougovski}}]{PhysRevLett.120.210501_prl_deuteron_8}
\bibinfo{author}{\bibfnamefont{E.~F.} \bibnamefont{Dumitrescu}},
  \bibinfo{author}{\bibfnamefont{A.~J.} \bibnamefont{McCaskey}},
  \bibinfo{author}{\bibfnamefont{G.}~\bibnamefont{Hagen}},
  \bibinfo{author}{\bibfnamefont{G.~R.} \bibnamefont{Jansen}},
  \bibinfo{author}{\bibfnamefont{T.~D.} \bibnamefont{Morris}},
  \bibinfo{author}{\bibfnamefont{T.}~\bibnamefont{Papenbrock}},
  \bibinfo{author}{\bibfnamefont{R.~C.} \bibnamefont{Pooser}},
  \bibinfo{author}{\bibfnamefont{D.~J.} \bibnamefont{Dean}}, \bibnamefont{and}
  \bibinfo{author}{\bibfnamefont{P.}~\bibnamefont{Lougovski}},
  \bibinfo{journal}{Phys. Rev. Lett.} \textbf{\bibinfo{volume}{120}},
  \bibinfo{pages}{210501} (\bibinfo{year}{2018}),
  \urlprefix\url{https://link.aps.org/doi/10.1103/PhysRevLett.120.210501}.

\bibitem[{\citenamefont{McArdle et~al.}(2019)\citenamefont{McArdle, Jones,
  Endo, Li, Benjamin, and Yuan}}]{variational_npj_21}
\bibinfo{author}{\bibfnamefont{S.}~\bibnamefont{McArdle}},
  \bibinfo{author}{\bibfnamefont{T.}~\bibnamefont{Jones}},
  \bibinfo{author}{\bibfnamefont{S.}~\bibnamefont{Endo}},
  \bibinfo{author}{\bibfnamefont{Y.}~\bibnamefont{Li}},
  \bibinfo{author}{\bibfnamefont{S.~C.} \bibnamefont{Benjamin}},
  \bibnamefont{and} \bibinfo{author}{\bibfnamefont{X.}~\bibnamefont{Yuan}},
  \bibinfo{journal}{npj Quantum Information} \textbf{\bibinfo{volume}{5}},
  \bibinfo{pages}{2056} (\bibinfo{year}{2019}),
  \urlprefix\url{https://doi.org/10.1038/s41534-019-0187-2}.

\bibitem[{\citenamefont{Turro et~al.}(2023)\citenamefont{Turro, Chistolini,
  Hashim, Kim, Livingston, Kreikebaum, Wendt, Dubois, Pederiva, Quaglioni
  et~al.}}]{PhysRevA.108.032417_reaction_3}
\bibinfo{author}{\bibfnamefont{F.}~\bibnamefont{Turro}},
  \bibinfo{author}{\bibfnamefont{T.}~\bibnamefont{Chistolini}},
  \bibinfo{author}{\bibfnamefont{A.}~\bibnamefont{Hashim}},
  \bibinfo{author}{\bibfnamefont{Y.}~\bibnamefont{Kim}},
  \bibinfo{author}{\bibfnamefont{W.}~\bibnamefont{Livingston}},
  \bibinfo{author}{\bibfnamefont{J.~M.} \bibnamefont{Kreikebaum}},
  \bibinfo{author}{\bibfnamefont{K.~A.} \bibnamefont{Wendt}},
  \bibinfo{author}{\bibfnamefont{J.~L.} \bibnamefont{Dubois}},
  \bibinfo{author}{\bibfnamefont{F.}~\bibnamefont{Pederiva}},
  \bibinfo{author}{\bibfnamefont{S.}~\bibnamefont{Quaglioni}},
  \bibnamefont{et~al.}, \bibinfo{journal}{Phys. Rev. A}
  \textbf{\bibinfo{volume}{108}}, \bibinfo{pages}{032417}
  (\bibinfo{year}{2023}),
  \urlprefix\url{https://link.aps.org/doi/10.1103/PhysRevA.108.032417}.

\bibitem[{\citenamefont{Yuan et~al.}(2019)\citenamefont{Yuan, Endo, Zhao, Li,
  and Benjamin}}]{Yuan2019theoryofvariational_12}
\bibinfo{author}{\bibfnamefont{X.}~\bibnamefont{Yuan}},
  \bibinfo{author}{\bibfnamefont{S.}~\bibnamefont{Endo}},
  \bibinfo{author}{\bibfnamefont{Q.}~\bibnamefont{Zhao}},
  \bibinfo{author}{\bibfnamefont{Y.}~\bibnamefont{Li}}, \bibnamefont{and}
  \bibinfo{author}{\bibfnamefont{S.~C.} \bibnamefont{Benjamin}},
  \bibinfo{journal}{{Quantum}} \textbf{\bibinfo{volume}{3}},
  \bibinfo{pages}{191} (\bibinfo{year}{2019}), ISSN \bibinfo{issn}{2521-327X},
  \urlprefix\url{https://doi.org/10.22331/q-2019-10-07-191}.

\bibitem[{\citenamefont{Sarma et~al.}(2023)\citenamefont{Sarma, Di~Matteo,
  Abhishek, and Srivastava}}]{PhysRevC.108.064305_chand}
\bibinfo{author}{\bibfnamefont{C.}~\bibnamefont{Sarma}},
  \bibinfo{author}{\bibfnamefont{O.}~\bibnamefont{Di~Matteo}},
  \bibinfo{author}{\bibfnamefont{A.}~\bibnamefont{Abhishek}}, \bibnamefont{and}
  \bibinfo{author}{\bibfnamefont{P.~C.} \bibnamefont{Srivastava}},
  \bibinfo{journal}{Phys. Rev. C} \textbf{\bibinfo{volume}{108}},
  \bibinfo{pages}{064305} (\bibinfo{year}{2023}),
  \urlprefix\url{https://link.aps.org/doi/10.1103/PhysRevC.108.064305}.

\bibitem[{\citenamefont{Cervia et~al.}(2021)\citenamefont{Cervia, Balantekin,
  Coppersmith, Johnson, Love, Poole, Robbins, and
  Saffman}}]{PhysRevC.104.024305_lipkin_model_11}
\bibinfo{author}{\bibfnamefont{M.~J.} \bibnamefont{Cervia}},
  \bibinfo{author}{\bibfnamefont{A.~B.} \bibnamefont{Balantekin}},
  \bibinfo{author}{\bibfnamefont{S.~N.} \bibnamefont{Coppersmith}},
  \bibinfo{author}{\bibfnamefont{C.~W.} \bibnamefont{Johnson}},
  \bibinfo{author}{\bibfnamefont{P.~J.} \bibnamefont{Love}},
  \bibinfo{author}{\bibfnamefont{C.}~\bibnamefont{Poole}},
  \bibinfo{author}{\bibfnamefont{K.}~\bibnamefont{Robbins}}, \bibnamefont{and}
  \bibinfo{author}{\bibfnamefont{M.}~\bibnamefont{Saffman}},
  \bibinfo{journal}{Phys. Rev. C} \textbf{\bibinfo{volume}{104}},
  \bibinfo{pages}{024305} (\bibinfo{year}{2021}),
  \urlprefix\url{https://link.aps.org/doi/10.1103/PhysRevC.104.024305}.

\bibitem[{\citenamefont{Stokes et~al.}(2020)\citenamefont{Stokes, Izaac,
  Killoran, and Carleo}}]{Stokes2020quantumnatural_14}
\bibinfo{author}{\bibfnamefont{J.}~\bibnamefont{Stokes}},
  \bibinfo{author}{\bibfnamefont{J.}~\bibnamefont{Izaac}},
  \bibinfo{author}{\bibfnamefont{N.}~\bibnamefont{Killoran}}, \bibnamefont{and}
  \bibinfo{author}{\bibfnamefont{G.}~\bibnamefont{Carleo}},
  \bibinfo{journal}{{Quantum}} \textbf{\bibinfo{volume}{4}},
  \bibinfo{pages}{269} (\bibinfo{year}{2020}), ISSN \bibinfo{issn}{2521-327X},
  \urlprefix\url{https://doi.org/10.22331/q-2020-05-25-269}.

\bibitem[{\citenamefont{Higgott et~al.}(2019)\citenamefont{Higgott, Wang, and
  Brierley}}]{Higgott2019variationalquantum_vqd_10}
\bibinfo{author}{\bibfnamefont{O.}~\bibnamefont{Higgott}},
  \bibinfo{author}{\bibfnamefont{D.}~\bibnamefont{Wang}}, \bibnamefont{and}
  \bibinfo{author}{\bibfnamefont{S.}~\bibnamefont{Brierley}},
  \bibinfo{journal}{{Quantum}} \textbf{\bibinfo{volume}{3}},
  \bibinfo{pages}{156} (\bibinfo{year}{2019}), ISSN \bibinfo{issn}{2521-327X},
  \urlprefix\url{https://doi.org/10.22331/q-2019-07-01-156}.

\bibitem[{\citenamefont{Ollitrault et~al.}(2020)\citenamefont{Ollitrault,
  Kandala, Chen, Barkoutsos, Mezzacapo, Pistoia, Sheldon, Woerner, Gambetta,
  and Tavernelli}}]{PhysRevResearch.2.043140_molecular_excitation_13}
\bibinfo{author}{\bibfnamefont{P.~J.} \bibnamefont{Ollitrault}},
  \bibinfo{author}{\bibfnamefont{A.}~\bibnamefont{Kandala}},
  \bibinfo{author}{\bibfnamefont{C.-F.} \bibnamefont{Chen}},
  \bibinfo{author}{\bibfnamefont{P.~K.} \bibnamefont{Barkoutsos}},
  \bibinfo{author}{\bibfnamefont{A.}~\bibnamefont{Mezzacapo}},
  \bibinfo{author}{\bibfnamefont{M.}~\bibnamefont{Pistoia}},
  \bibinfo{author}{\bibfnamefont{S.}~\bibnamefont{Sheldon}},
  \bibinfo{author}{\bibfnamefont{S.}~\bibnamefont{Woerner}},
  \bibinfo{author}{\bibfnamefont{J.~M.} \bibnamefont{Gambetta}},
  \bibnamefont{and}
  \bibinfo{author}{\bibfnamefont{I.}~\bibnamefont{Tavernelli}},
  \bibinfo{journal}{Phys. Rev. Res.} \textbf{\bibinfo{volume}{2}},
  \bibinfo{pages}{043140} (\bibinfo{year}{2020}),
  \urlprefix\url{https://link.aps.org/doi/10.1103/PhysRevResearch.2.043140}.

\bibitem[{\citenamefont{Tang et~al.}(2021)\citenamefont{Tang, Shkolnikov,
  Barron, Grimsley, Mayhall, Barnes, and
  Economou}}]{PRXQuantum.2.020310_qubit_adapt_vqe_15}
\bibinfo{author}{\bibfnamefont{H.~L.} \bibnamefont{Tang}},
  \bibinfo{author}{\bibfnamefont{V.}~\bibnamefont{Shkolnikov}},
  \bibinfo{author}{\bibfnamefont{G.~S.} \bibnamefont{Barron}},
  \bibinfo{author}{\bibfnamefont{H.~R.} \bibnamefont{Grimsley}},
  \bibinfo{author}{\bibfnamefont{N.~J.} \bibnamefont{Mayhall}},
  \bibinfo{author}{\bibfnamefont{E.}~\bibnamefont{Barnes}}, \bibnamefont{and}
  \bibinfo{author}{\bibfnamefont{S.~E.} \bibnamefont{Economou}},
  \bibinfo{journal}{PRX Quantum} \textbf{\bibinfo{volume}{2}},
  \bibinfo{pages}{020310} (\bibinfo{year}{2021}),
  \urlprefix\url{https://link.aps.org/doi/10.1103/PRXQuantum.2.020310}.

\bibitem[{\citenamefont{Romero et~al.}(2022)\citenamefont{Romero, Engel, Tang,
  and Economou}}]{PhysRevC.105.064317_adapt_vqe_16}
\bibinfo{author}{\bibfnamefont{A.~M.} \bibnamefont{Romero}},
  \bibinfo{author}{\bibfnamefont{J.}~\bibnamefont{Engel}},
  \bibinfo{author}{\bibfnamefont{H.~L.} \bibnamefont{Tang}}, \bibnamefont{and}
  \bibinfo{author}{\bibfnamefont{S.~E.} \bibnamefont{Economou}},
  \bibinfo{journal}{Phys. Rev. C} \textbf{\bibinfo{volume}{105}},
  \bibinfo{pages}{064317} (\bibinfo{year}{2022}),
  \urlprefix\url{https://link.aps.org/doi/10.1103/PhysRevC.105.064317}.

\bibitem[{\citenamefont{Grimsley et~al.}(2019)\citenamefont{Grimsley, Economou,
  Barnes, and Mayhall}}]{adapt_nature_22}
\bibinfo{author}{\bibfnamefont{H.~R.} \bibnamefont{Grimsley}},
  \bibinfo{author}{\bibfnamefont{S.~E.} \bibnamefont{Economou}},
  \bibinfo{author}{\bibfnamefont{E.}~\bibnamefont{Barnes}}, \bibnamefont{and}
  \bibinfo{author}{\bibfnamefont{N.~J.} \bibnamefont{Mayhall}},
  \bibinfo{journal}{Nature Communications} \textbf{\bibinfo{volume}{10}},
  \bibinfo{pages}{2041} (\bibinfo{year}{2019}),
  \urlprefix\url{https://doi.org/10.1038/s41467-019-10988-2}.

\bibitem[{\citenamefont{Xie et~al.}(2024)\citenamefont{Xie, Xue, and
  Zhang}}]{xie:fp2024}
\bibinfo{author}{\bibfnamefont{X.-D.} \bibnamefont{Xie}},
  \bibinfo{author}{\bibfnamefont{Z.-Y.} \bibnamefont{Xue}}, \bibnamefont{and}
  \bibinfo{author}{\bibfnamefont{D.-B.} \bibnamefont{Zhang}},
  \bibinfo{journal}{Frontiers of Physics} \textbf{\bibinfo{volume}{19}},
  \bibinfo{pages}{41202} (\bibinfo{year}{2024}),
  \urlprefix\url{https://doi.org/10.1007/s11467-023-1382-3}.

\bibitem[{\citenamefont{Bian and Kais}(2021)}]{Bian:jcp2021}
\bibinfo{author}{\bibfnamefont{T.}~\bibnamefont{Bian}} \bibnamefont{and}
  \bibinfo{author}{\bibfnamefont{S.}~\bibnamefont{Kais}}, \bibinfo{journal}{The
  Journal of Chemical Physics} \textbf{\bibinfo{volume}{154}},
  \bibinfo{pages}{194107} (\bibinfo{year}{2021}), ISSN
  \bibinfo{issn}{0021-9606}, \urlprefix\url{https://doi.org/10.1063/5.0040477}.

\bibitem[{\citenamefont{Zhang et~al.}(2024)\citenamefont{Zhang, Bai, and
  Ren}}]{zhang:plb2024}
\bibinfo{author}{\bibfnamefont{H.}~\bibnamefont{Zhang}},
  \bibinfo{author}{\bibfnamefont{D.}~\bibnamefont{Bai}}, \bibnamefont{and}
  \bibinfo{author}{\bibfnamefont{Z.}~\bibnamefont{Ren}},
  \bibinfo{journal}{Physics Letters B} p. \bibinfo{pages}{139187}
  (\bibinfo{year}{2024}), ISSN \bibinfo{issn}{0370-2693},
  \urlprefix\url{https://www.sciencedirect.com/science/article/pii/S0370269324007457}.

\bibitem[{\citenamefont{Zhao et~al.}(2023)\citenamefont{Zhao, Zhang, and
  Wei}}]{Zhao:sr2023}
\bibinfo{author}{\bibfnamefont{H.}~\bibnamefont{Zhao}},
  \bibinfo{author}{\bibfnamefont{P.}~\bibnamefont{Zhang}}, \bibnamefont{and}
  \bibinfo{author}{\bibfnamefont{T.-C.} \bibnamefont{Wei}},
  \bibinfo{journal}{Sci Rep} \textbf{\bibinfo{volume}{13}},
  \bibinfo{pages}{22313} (\bibinfo{year}{2023}),
  \urlprefix\url{https://doi.org/10.1038/s41598-023-49662-5}.

\bibitem[{\citenamefont{Myo and Katō}(2020)}]{myo:2020ptepptaa101}
\bibinfo{author}{\bibfnamefont{T.}~\bibnamefont{Myo}} \bibnamefont{and}
  \bibinfo{author}{\bibfnamefont{K.}~\bibnamefont{Katō}},
  \bibinfo{journal}{Progress of Theoretical and Experimental Physics}
  \textbf{\bibinfo{volume}{2020}} (\bibinfo{year}{2020}), ISSN
  \bibinfo{issn}{2050-3911}, \bibinfo{note}{12A101},
  \eprint{https://academic.oup.com/ptep/article-pdf/2020/12/12A101/35612142/ptaa101.pdf},
  \urlprefix\url{https://doi.org/10.1093/ptep/ptaa101}.

\bibitem[{\citenamefont{Aguilar and Combes}(1971)}]{abc:Aguilar}
\bibinfo{author}{\bibfnamefont{J.}~\bibnamefont{Aguilar}} \bibnamefont{and}
  \bibinfo{author}{\bibfnamefont{J.~M.} \bibnamefont{Combes}},
  \bibinfo{journal}{Communications in Mathematical Physics}
  \textbf{\bibinfo{volume}{22}}, \bibinfo{pages}{269} (\bibinfo{year}{1971}),
  ISSN \bibinfo{issn}{1432-0916},
  \urlprefix\url{https://doi.org/10.1007/BF01877510}.

\bibitem[{\citenamefont{Balslev and Combes}(1971)}]{abc:Balsev}
\bibinfo{author}{\bibfnamefont{E.}~\bibnamefont{Balslev}} \bibnamefont{and}
  \bibinfo{author}{\bibfnamefont{J.~M.} \bibnamefont{Combes}},
  \bibinfo{journal}{Communications in Mathematical Physics}
  \textbf{\bibinfo{volume}{22}}, \bibinfo{pages}{280} (\bibinfo{year}{1971}),
  ISSN \bibinfo{issn}{1432-0916},
  \urlprefix\url{https://doi.org/10.1007/BF01877511}.

\bibitem[{\citenamefont{Kruppa et~al.}(1988)\citenamefont{Kruppa, Lovas, and
  Gyarmati}}]{Moiseyev:PhysRevC.37.383}
\bibinfo{author}{\bibfnamefont{A.~T.} \bibnamefont{Kruppa}},
  \bibinfo{author}{\bibfnamefont{R.~G.} \bibnamefont{Lovas}}, \bibnamefont{and}
  \bibinfo{author}{\bibfnamefont{B.}~\bibnamefont{Gyarmati}},
  \bibinfo{journal}{Phys. Rev. C} \textbf{\bibinfo{volume}{37}},
  \bibinfo{pages}{383} (\bibinfo{year}{1988}),
  \urlprefix\url{https://link.aps.org/doi/10.1103/PhysRevC.37.383}.

\bibitem[{\citenamefont{{Moiseyev}}(2011)}]{2011nhqm.book.....M}
\bibinfo{author}{\bibfnamefont{N.}~\bibnamefont{{Moiseyev}}},
  \emph{\bibinfo{title}{{Non-Hermitian Quantum Mechanics}}}
  (\bibinfo{publisher}{{Cambridge University Press}}, \bibinfo{year}{2011}).

\bibitem[{\citenamefont{{Kraf}}(2013)}]{d_kraft_2013_thessis}
\bibinfo{author}{\bibfnamefont{D.}~\bibnamefont{{Kraf}}},
  \emph{\bibinfo{title}{{Stochastic Variational Approaches to Non-Hermitian
  Quantum-Mechanical Problems}}} (\bibinfo{year}{2013}).

\bibitem[{\citenamefont{Hiyama et~al.}(2003)\citenamefont{Hiyama, Kino, and
  Kamimura}}]{HIYAMA:ppnp2003}
\bibinfo{author}{\bibfnamefont{E.}~\bibnamefont{Hiyama}},
  \bibinfo{author}{\bibfnamefont{Y.}~\bibnamefont{Kino}}, \bibnamefont{and}
  \bibinfo{author}{\bibfnamefont{M.}~\bibnamefont{Kamimura}},
  \bibinfo{journal}{Progress in Particle and Nuclear Physics}
  \textbf{\bibinfo{volume}{51}}, \bibinfo{pages}{223} (\bibinfo{year}{2003}),
  ISSN \bibinfo{issn}{0146-6410},
  \urlprefix\url{https://www.sciencedirect.com/science/article/pii/S0146641003900159}.

\bibitem[{\citenamefont{Cs\'ot\'o et~al.}(1990)\citenamefont{Cs\'ot\'o,
  Gyarmati, Kruppa, P\'al, and Moiseyev}}]{csoto:pra1990}
\bibinfo{author}{\bibfnamefont{A.}~\bibnamefont{Cs\'ot\'o}},
  \bibinfo{author}{\bibfnamefont{B.}~\bibnamefont{Gyarmati}},
  \bibinfo{author}{\bibfnamefont{A.~T.} \bibnamefont{Kruppa}},
  \bibinfo{author}{\bibfnamefont{K.~F.} \bibnamefont{P\'al}}, \bibnamefont{and}
  \bibinfo{author}{\bibfnamefont{N.}~\bibnamefont{Moiseyev}},
  \bibinfo{journal}{Phys. Rev. A} \textbf{\bibinfo{volume}{41}},
  \bibinfo{pages}{3469} (\bibinfo{year}{1990}),
  \urlprefix\url{https://link.aps.org/doi/10.1103/PhysRevA.41.3469}.

\bibitem[{\citenamefont{Moiseyev et~al.}(1978)\citenamefont{Moiseyev, Certain,
  and Weinhold}}]{Moiseyev:mole1978}
\bibinfo{author}{\bibfnamefont{N.}~\bibnamefont{Moiseyev}},
  \bibinfo{author}{\bibfnamefont{P.~R.} \bibnamefont{Certain}},
  \bibnamefont{and} \bibinfo{author}{\bibfnamefont{F.}~\bibnamefont{Weinhold}},
  \bibinfo{journal}{Molecular Physics} \textbf{\bibinfo{volume}{36}},
  \bibinfo{pages}{1613} (\bibinfo{year}{1978}), ISSN \bibinfo{issn}{1362-3028},
  \urlprefix\url{https://doi.org/10.1080/00268977800102631}.

\bibitem[{\citenamefont{Rittby et~al.}(1981)\citenamefont{Rittby, Elander, and
  Br\"andas}}]{Rittby:PhysRevA.24.1636}
\bibinfo{author}{\bibfnamefont{M.}~\bibnamefont{Rittby}},
  \bibinfo{author}{\bibfnamefont{N.}~\bibnamefont{Elander}}, \bibnamefont{and}
  \bibinfo{author}{\bibfnamefont{E.}~\bibnamefont{Br\"andas}},
  \bibinfo{journal}{Phys. Rev. A} \textbf{\bibinfo{volume}{24}},
  \bibinfo{pages}{1636} (\bibinfo{year}{1981}),
  \urlprefix\url{https://link.aps.org/doi/10.1103/PhysRevA.24.1636}.

\bibitem[{\citenamefont{Buck et~al.}(1977)\citenamefont{Buck, Friedrich, and
  Wheatley}}]{Buck:npa1977}
\bibinfo{author}{\bibfnamefont{B.}~\bibnamefont{Buck}},
  \bibinfo{author}{\bibfnamefont{H.}~\bibnamefont{Friedrich}},
  \bibnamefont{and} \bibinfo{author}{\bibfnamefont{C.}~\bibnamefont{Wheatley}},
  \bibinfo{journal}{Nuclear Physics A} \textbf{\bibinfo{volume}{275}},
  \bibinfo{pages}{246} (\bibinfo{year}{1977}), ISSN \bibinfo{issn}{0375-9474},
  \urlprefix\url{https://www.sciencedirect.com/science/article/pii/0375947477902871}.

\bibitem[{\citenamefont{Seeley et~al.}(2012)\citenamefont{Seeley, Richard, and
  Love}}]{10.1063/1.4768229_bk_chem}
\bibinfo{author}{\bibfnamefont{J.~T.} \bibnamefont{Seeley}},
  \bibinfo{author}{\bibfnamefont{M.~J.} \bibnamefont{Richard}},
  \bibnamefont{and} \bibinfo{author}{\bibfnamefont{P.~J.} \bibnamefont{Love}},
  \bibinfo{journal}{The Journal of Chemical Physics}
  \textbf{\bibinfo{volume}{137}}, \bibinfo{pages}{224109}
  (\bibinfo{year}{2012}), ISSN \bibinfo{issn}{0021-9606},
  \eprint{https://pubs.aip.org/aip/jcp/article-pdf/doi/10.1063/1.4768229/13999577/224109\_1\_online.pdf},
  \urlprefix\url{https://doi.org/10.1063/1.4768229}.

\bibitem[{\citenamefont{Bravyi and Kitaev}(2002)}]{BRAVYI2002210_bk}
\bibinfo{author}{\bibfnamefont{S.~B.} \bibnamefont{Bravyi}} \bibnamefont{and}
  \bibinfo{author}{\bibfnamefont{A.~Y.} \bibnamefont{Kitaev}},
  \bibinfo{journal}{Annals of Physics} \textbf{\bibinfo{volume}{298}},
  \bibinfo{pages}{210} (\bibinfo{year}{2002}), ISSN \bibinfo{issn}{0003-4916},
  \urlprefix\url{https://www.sciencedirect.com/science/article/pii/S0003491602962548}.

\bibitem[{\citenamefont{Di~Matteo et~al.}(2021)\citenamefont{Di~Matteo, McCoy,
  Gysbers, Miyagi, Woloshyn, and Navr\'atil}}]{improve_gc_olivia}
\bibinfo{author}{\bibfnamefont{O.}~\bibnamefont{Di~Matteo}},
  \bibinfo{author}{\bibfnamefont{A.}~\bibnamefont{McCoy}},
  \bibinfo{author}{\bibfnamefont{P.}~\bibnamefont{Gysbers}},
  \bibinfo{author}{\bibfnamefont{T.}~\bibnamefont{Miyagi}},
  \bibinfo{author}{\bibfnamefont{R.~M.} \bibnamefont{Woloshyn}},
  \bibnamefont{and}
  \bibinfo{author}{\bibfnamefont{P.}~\bibnamefont{Navr\'atil}},
  \bibinfo{journal}{Phys. Rev. A} \textbf{\bibinfo{volume}{103}},
  \bibinfo{pages}{042405} (\bibinfo{year}{2021}),
  \urlprefix\url{https://link.aps.org/doi/10.1103/PhysRevA.103.042405}.

\bibitem[{\citenamefont{Siwach and
  Arumugam}(2021)}]{pooja_PhysRevC.104.034301_gc}
\bibinfo{author}{\bibfnamefont{P.}~\bibnamefont{Siwach}} \bibnamefont{and}
  \bibinfo{author}{\bibfnamefont{P.}~\bibnamefont{Arumugam}},
  \bibinfo{journal}{Phys. Rev. C} \textbf{\bibinfo{volume}{104}},
  \bibinfo{pages}{034301} (\bibinfo{year}{2021}),
  \urlprefix\url{https://link.aps.org/doi/10.1103/PhysRevC.104.034301}.

\bibitem[{\citenamefont{Jordan and Wigner}(2024)}]{jw}
\bibinfo{author}{\bibfnamefont{P.}~\bibnamefont{Jordan}} \bibnamefont{and}
  \bibinfo{author}{\bibfnamefont{E.}~\bibnamefont{Wigner}},
  \bibinfo{journal}{Zeitschrift für Physik} \textbf{\bibinfo{volume}{47}},
  \bibinfo{pages}{631} (\bibinfo{year}{2024}),
  \urlprefix\url{https://doi.org/10.1007/BF01331938}.

\bibitem[{\citenamefont{Lacroix}(2020)}]{Lacroix:prl2020}
\bibinfo{author}{\bibfnamefont{D.}~\bibnamefont{Lacroix}},
  \bibinfo{journal}{Phys. Rev. Lett.} \textbf{\bibinfo{volume}{125}},
  \bibinfo{pages}{230502} (\bibinfo{year}{2020}),
  \urlprefix\url{https://link.aps.org/doi/10.1103/PhysRevLett.125.230502}.

\bibitem[{\citenamefont{Siwach and Lacroix}(2021)}]{Siwach:pra2021}
\bibinfo{author}{\bibfnamefont{P.}~\bibnamefont{Siwach}} \bibnamefont{and}
  \bibinfo{author}{\bibfnamefont{D.}~\bibnamefont{Lacroix}},
  \bibinfo{journal}{Phys. Rev. A} \textbf{\bibinfo{volume}{104}},
  \bibinfo{pages}{062435} (\bibinfo{year}{2021}),
  \urlprefix\url{https://link.aps.org/doi/10.1103/PhysRevA.104.062435}.

\bibitem[{\citenamefont{Homma et~al.}(1997)\citenamefont{Homma, Myo, and
  Katō}}]{mayo:1997ptp}
\bibinfo{author}{\bibfnamefont{M.}~\bibnamefont{Homma}},
  \bibinfo{author}{\bibfnamefont{T.}~\bibnamefont{Myo}}, \bibnamefont{and}
  \bibinfo{author}{\bibfnamefont{K.}~\bibnamefont{Katō}},
  \bibinfo{journal}{Progress of Theoretical Physics}
  \textbf{\bibinfo{volume}{97}}, \bibinfo{pages}{561} (\bibinfo{year}{1997}),
  ISSN \bibinfo{issn}{0033-068X},
  \urlprefix\url{https://doi.org/10.1143/PTP.97.561}.

\bibitem[{\citenamefont{Javadi-Abhari et~al.}(2024)\citenamefont{Javadi-Abhari,
  Treinish, Krsulich, Wood, Lishman, Gacon, Martiel, Nation, Bishop, Cross
  et~al.}}]{qiskit2024}
\bibinfo{author}{\bibfnamefont{A.}~\bibnamefont{Javadi-Abhari}},
  \bibinfo{author}{\bibfnamefont{M.}~\bibnamefont{Treinish}},
  \bibinfo{author}{\bibfnamefont{K.}~\bibnamefont{Krsulich}},
  \bibinfo{author}{\bibfnamefont{C.~J.} \bibnamefont{Wood}},
  \bibinfo{author}{\bibfnamefont{J.}~\bibnamefont{Lishman}},
  \bibinfo{author}{\bibfnamefont{J.}~\bibnamefont{Gacon}},
  \bibinfo{author}{\bibfnamefont{S.}~\bibnamefont{Martiel}},
  \bibinfo{author}{\bibfnamefont{P.~D.} \bibnamefont{Nation}},
  \bibinfo{author}{\bibfnamefont{L.~S.} \bibnamefont{Bishop}},
  \bibinfo{author}{\bibfnamefont{A.~W.} \bibnamefont{Cross}},
  \bibnamefont{et~al.}, \emph{\bibinfo{title}{Quantum computing with {Q}iskit}}
  (\bibinfo{year}{2024}), \eprint{2405.08810}.

\bibitem[{\citenamefont{Virtanen et~al.}(2020)\citenamefont{Virtanen, Gommers,
  Oliphant, Haberland, Reddy, Cournapeau, Burovski, Peterson, Weckesser, Bright
  et~al.}}]{2020SciPy-NMeth}
\bibinfo{author}{\bibfnamefont{P.}~\bibnamefont{Virtanen}},
  \bibinfo{author}{\bibfnamefont{R.}~\bibnamefont{Gommers}},
  \bibinfo{author}{\bibfnamefont{T.~E.} \bibnamefont{Oliphant}},
  \bibinfo{author}{\bibfnamefont{M.}~\bibnamefont{Haberland}},
  \bibinfo{author}{\bibfnamefont{T.}~\bibnamefont{Reddy}},
  \bibinfo{author}{\bibfnamefont{D.}~\bibnamefont{Cournapeau}},
  \bibinfo{author}{\bibfnamefont{E.}~\bibnamefont{Burovski}},
  \bibinfo{author}{\bibfnamefont{P.}~\bibnamefont{Peterson}},
  \bibinfo{author}{\bibfnamefont{W.}~\bibnamefont{Weckesser}},
  \bibinfo{author}{\bibfnamefont{J.}~\bibnamefont{Bright}},
  \bibnamefont{et~al.}, \bibinfo{journal}{Nature Methods}
  \textbf{\bibinfo{volume}{17}}, \bibinfo{pages}{261} (\bibinfo{year}{2020}).

\end{thebibliography}
\end{document}